\documentclass[fleqn,usenatbib]{mnras}

\usepackage[T1]{fontenc}
\usepackage{ae,aecompl}
\usepackage{graphicx}	% Including Figure files
\usepackage{amsmath}	% Advanced maths commands
\usepackage{amssymb}	% Extra maths symbols
\usepackage{booktabs}
\usepackage[table,xcdraw]{xcolor}
\usepackage[T1]{fontenc}
\usepackage{ae,aecompl}
\usepackage{marvosym}
\usepackage{hyperref}
\hypersetup{colorlinks,allcolors=cyan}
\usepackage{soul}
\usepackage{booktabs}
\usepackage{multirow}
\usepackage{caption}
\usepackage{subcaption}
\usepackage{comment}
\usepackage{ulem}
\usepackage{flushend}
\usepackage{newtxtext,newtxmath}

\newcommand{\beq}[1]{\begin{equation}\label{#1}}
\newcommand{\eeq}{\end{equation}}
\newcommand{\sub}[1]{_{\rm #1}}
\newcommand{\beqn}{\begin{equation}}
\newcommand{\eeqn}{\end{equation}}
\newcommand{\Op}{\Omega\sub{p}}

\newcommand{\aroche}{a\sub{Roche}}

\newcommand{\Mm}{M\sub{m}}

\newcommand{\Mmo}{M\sub{m,o}}
\newcommand{\Rmo}{R_\mathrm{m,o}}
\newcommand{\amo}{a\sub{m,o}}

\newcommand{\Rmt}{R_\mathrm{m,t}}

\newcommand{\ap}{a\sub{p}}
\newcommand{\astop}{a\sub{\mathrm{stop}}}

\newcommand{\Rs}{R_\star}
\newcommand{\Rm}{R_\mathrm{m}}

\newcommand{\Mp}{M\sub{p}}
\newcommand{\Rh}{R\sub{Hill}}

\newcommand{\Rp}{R\sub{p}}
\newcommand{\Pp}{P\sub{p}}
\newcommand{\Pm}{P\sub{m}}

\newcommand{\iR}{i\sub{R}}

\newcommand{\Rint}{r\sub{i}}
\newcommand{\Rou}{r\sub{o}}

\newcommand{\tpr}{t_\mathrm{PR}}
\newcommand{\Qpr}{Q\sub{PR}}

\newcommand{\Der}{\mathrm{d}}

\newcommand{\SigmaSB}{\sigma\sub{SB}}

\newcommand{\Kpp}{k\sub{2}}
\newcommand{\Qp}{Q}

\newcommand{\Mmoon}{M\sub{m}}
\newcommand{\amoon}{a\sub{m}}
\newcommand{\nmm}{n\sub{m}}
\def\sgn{\mathrm{sgn}\,}
\newcommand{\Mstar}{M\sub{\star}}
\newcommand{\apos}{a\sub{p}}
\newcommand{\npp}{n\sub{p}}

\newcommand{\Teq}{T\sub{eq}}

\newcommand{\Lstar}{L\sub{\star}}

\definecolor{mygreen}{RGB}{150,1,122}

\newcommand{\rev}[1]{\textcolor{black}{ #1}}

\usepackage{comment}

\title[Cronomoons]{{\it Cronomoons}: origin, dynamics, and light-curve features of ringed exomoons} % 

\author[Sucerquia et al.]{\parbox{\textwidth}{
Mario Sucerquia,$^{1,2}$\thanks{E-mail:\href{mailto:mario.sucerquia@uv.cl}{mario.sucerquia@uv.cl}} Jaime A. Alvarado-Montes,$^{3, 4}$ Amelia Bayo,$^{1,2}$ Jorge Cuadra,$^{5,2}$ Nicol\'as Cuello,$^{6}$ Cristian A. Giuppone,$^{7}$ Mat\'ias Montesinos,$^{8, 2}$ J. Olofsson,$^{1,2,9}$ Christian Schwab,$^{3,4}$ Lee Spitler,$^{3,4,10}$ and Jorge I. Zuluaga$^{11}$.
}\vspace{0.4cm} \\
$^{1}$Instituto de F\'isica y Astronom\'ia, Facultad de Ciencias, Universidad de Valpara\'iso, Av. Gran Breta\~na 1111, 5030 Casilla, Valpara\'iso, Chile\\
$^{2}$N\'ucleo Milenio Formaci\'on Planetaria - NPF, Universidad de Valpara\'iso, Av. Gran Breta\~na 1111, Valpara\'iso, Chile \\
$^{3}$Department of Physics \& Astronomy, Macquarie University -- Sydney, NSW 2109, Australia.\\
$^{4}$Centre for Astronomy, Astrophysics and Astrophotonics, Macquarie University -- Sydney, NSW 2109, Australia.\\
$^{5}$Departamento de Ciencias, Facultad de Artes Liberales, Universidad Adolfo Ib\'a\~nez, Avenida Padre Hurtado 750, Vi\~na del Mar, Chile.\\
$^{6}$Univ. Grenoble Alpes, CNRS, IPAG, F-38000 Grenoble, France.\\
$^{7}$Universidad Nacional de C\'ordoba, Observatorio Astron\'omico - IATE. Laprida 854, 5000 C\'ordoba, Argentina\\
$^{8}$Escuela de Ciencias, Universidad Vi\~na del Mar, Vi\~na
del Mar, Chile.\\
$^{9}$Max Planck Institute for Astronomy, Königstuhl 17, D-69117 Heidelberg, Germany\\
$^{10}$Australian Astronomical Optics, 105 Delhi Rd, North Ryde, NSW 2113, Australia\\
$^{11}$SEAP research group, Instituto de F\'{\i}sica, FCEN, Universidad de Antioquia -- Calle 70 No. 52-21, Medell\'in, Colombia.
}

\date{Accepted 30 November 2021. Received 03 August 2021; in original form 22 July 2021}

\pubyear{2021}

\begin{document}
\label{firstpage}
\pagerange{\pageref{firstpage}--\pageref{lastpage}}
\maketitle

% Abstract of the paper
\begin{abstract}
In recent years, technical and theoretical work to detect moons and rings around exoplanets has been attempted. The small mass/size ratios between moons and planets means this is very challenging, having only one exoplanetary system where spotting an exomoon might be feasible (i.e. Kepler-1625b i). In this work, we study the dynamical evolution of ringed exomoons, dubbed {\it cronomoons} after their similarity with Cronus (Greek for Saturn), and after Chronos (the epitome of time), following the Transit Timing Variations (TTV) and Transit Duration Variation (TDV) that they produce on their host planet. \textit{Cronomoons} have extended systems of rings that make them appear bigger than they actually are when transiting in front of their host star. We explore different possible scenarios that could lead to the formation of such circumsatellital rings, and through the study of the dynamical/thermodynamic stability and lifespan of their dust and ice ring particles, we found that an isolated \textit{cronomoon} can survive for time-scales long enough to be detected and followed up. If these objects exist, {\it cronomoons}' rings will exhibit gaps similar to Saturn's Cassini Division and analogous to the asteroid belt's {\it Kirkwood gaps}, but instead raised due to resonances induced by the host planet. Finally, we analyse the case of Kepler-1625b i under the scope of this work, finding that the controversial giant moon could instead be an Earth-mass \textit{cronomoon}. From a theoretical perspective, this scenario can contribute to a better interpretation of the underlying phenomenology in current and future observations.
\end{abstract}

% Select between one and six entries from the list of approved keywords.
% Don't make up new ones.

%%%%%%%%%%%%%%%%%%%%%%%%%%%%%%%%%%%%%%%%%%%%%%%%%%

%%%%%%%%%%%%%%%%% BODY OF PAPER %%%%%%%%%%%%%%%%%%

\section{Introduction}
%FFFFFFFFFFFFFFFFFFFFFFFFFFFFFFFFFF
%FFFFFFFFFFFFFFFFFFFFFFFFFFFFFFFFFF
\begin{figure*}
	\includegraphics[width=\textwidth]{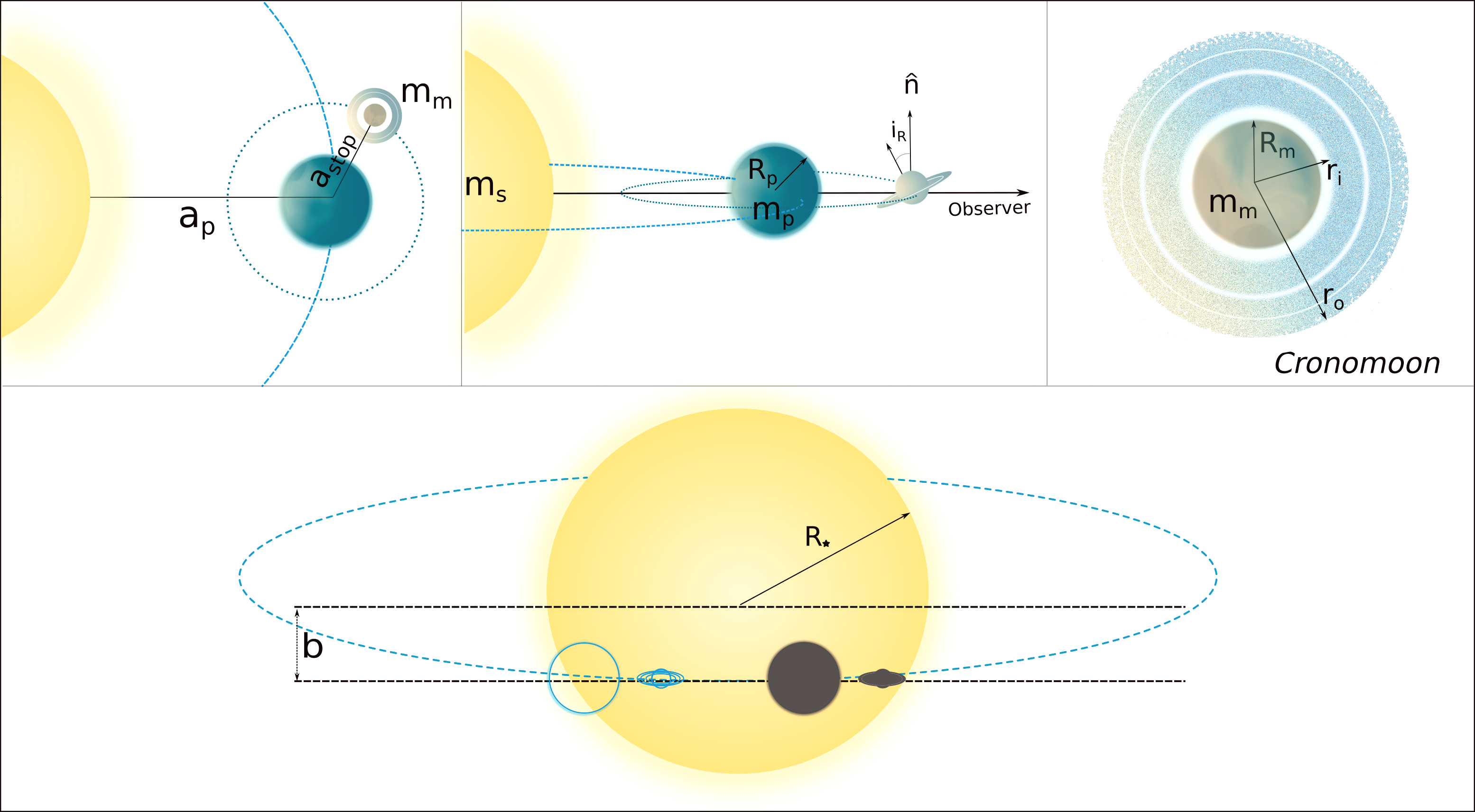}
    \caption{Schematic illustration of the system considered in our model. We highlight the semi-major axis of the planet ($\ap$) and the moon's ($\astop$); the mass and radius of the star, the planet, and the moon ($\Mstar$, $\Mp$, and $\Mm$; $\Rs$, $\Rp$ and $\Rm$, respectively). The ring's tilt, its inner and outer radii are also shown ($\iR, \Rm, \Rint$ and $\Rou$), as well as the transit impact parameter $b$.}
    \label{fig:scheme}
\end{figure*}
%FFFFFFFFFFFFFFFFFFFFFFFFFFFFFFFFFF
%FFFFFFFFFFFFFFFFFFFFFFFFFFFFFFFFFF

The first confirmation of \rev{an exomoon} would certainly be \rev{an important step to refine both observational and theoretical predictions of exoplanetary systems. Finding an ensemble of moons would enable a new tool, giving us} more leverage to better understand the formation and dynamical evolution of planetary systems. According to formation models, moons should be common among Jupiter-like planets discovered within the circumstellar habitable zones of their host stars (see e.g. \citealt{Heller2013astrob,Heller2014,Martinez2019}). As this class of exoplanet is known to be numerous, a successful confirmation (or lack thereof) will have strong implications for models of migration and habitability of moons.

\rev{There were recently six spurious claims for exomoon candidates by \citet{Fox2021}, which have been proven incorrect from both an observational \citep{Kipping2020} and a stability perspective \citep{Quarles2021}}. Does this reflect a lack of moons? or is it insufficient sensitivity of current observations? Direct detection of such satellites could be achievable for the next generation of telescopes like the European Extremely Large Telescope \citep[E-ELTS;][]{Peters2013}, whereas radial velocity and astrometric methods are probably incapable of reaching these targets \citep[][]{Vanderburg2018}. Perhaps the only feasible way is the analysis of light curve data in systems with multiple transits of planets orbited by large exomoons. This would involve the detection of one or more second-order dynamical effects on planetary light curves: Transit Duration Variations (TDVs), Transit Timing Variations \citep[TTVs;][]{Kipping2012,Kipping2013ApJ,Kipping2014ApJ}, and Transit Radius Variations \citep*[TRVs;][]{Rodenbeck2020}.

The aforementioned effects (TDVs, TTVs, TRVs) could be useful to assess the physical and orbital properties of exoplanets and exomoons. So far, these are the most promising techniques to reveal the presence of (yet undetected) low-mass companions such as rocky planets, rings, moons, and even smaller objects (see e.g. \citealt{Seager03,Kipping2009a, Zuluaga2015, Sucerquia2020, Sucerquia2020b}, and references therein). TTVs are related to the changes in the planet's orbital position at the beginning of its transit which could reveal the presence of nearby planets and/or satellites; while TDVs are variations in the duration of transits, emerging as a consequence of changes in the velocity of the planet during the transit, produced by the presence of a companion satellite. Finally, TRVs are due to small time-dependent contributions of moons to the transit depth, which happens during a transit when moons lie in the line of sight of the observer. Further information about the presence of rings, cometary tails, or other circumplanetary structures could be encoded in the photometric shape of the ingress and egress stages of planetary transits (see e.g, \citealt{Zuluaga2015,Rappaport2018}; \citealt*{Arkhypov2021}).

Kepler-1625b i \citep{Teachey-Kipping-2018} is currently the most convincing exomoon candidate spotted by the detection of a mutual transit, \rev{and by re-drawing a new linear ephemeris through all data points as of yet, \citet{Kipping2021} has calculated a TTV of $\sim 17$ min for this system}. However, interpretation of this observational result is still being debated (see e.g.
\citealt*{Heller2019};
\citealt*{Kreidberg2019}; \citealt{Teachey_2020}). The main uncertainty in interpretation lies in the planet--moon physical and orbital characteristics estimated for this system: the mass-ratio $\Mmoon/\Mp \sim 10^{-3 \, \text{to} \, -2 }$ (where $M_m$ and $M_p$ are the mass of the moon and the planet, respectively) in particular suggests that Kepler-1625b i is unlikely to have formed in the Galilean moon formation framework \citep{Canup2006}.

The aim of this work is to study the dynamical stability of a gravitationally bound system composed of a single-planet being orbited by a ringed moon (see Fig. \ref{fig:scheme}). We name these objects \textit{cronomoons} after the planet Saturn (Cronus, in Greek mythology) and Chronos (the epitome of time) in relation to the set of unique observational signatures that this kind of system might exhibit, namely large TRVs but small TTVs and TDVs. However, this situation could also occur, for example, in exoplanet systems where moons have large, dense, and thick atmospheres sufficient to obscure a large portion of the stellar disc \citep{Rodenbeck2020}, but with considerably small bulk masses unable to cause detectable perturbations on the TTV and TDV signals.

\begin{figure*}
     \centering
        \includegraphics[width=\textwidth]{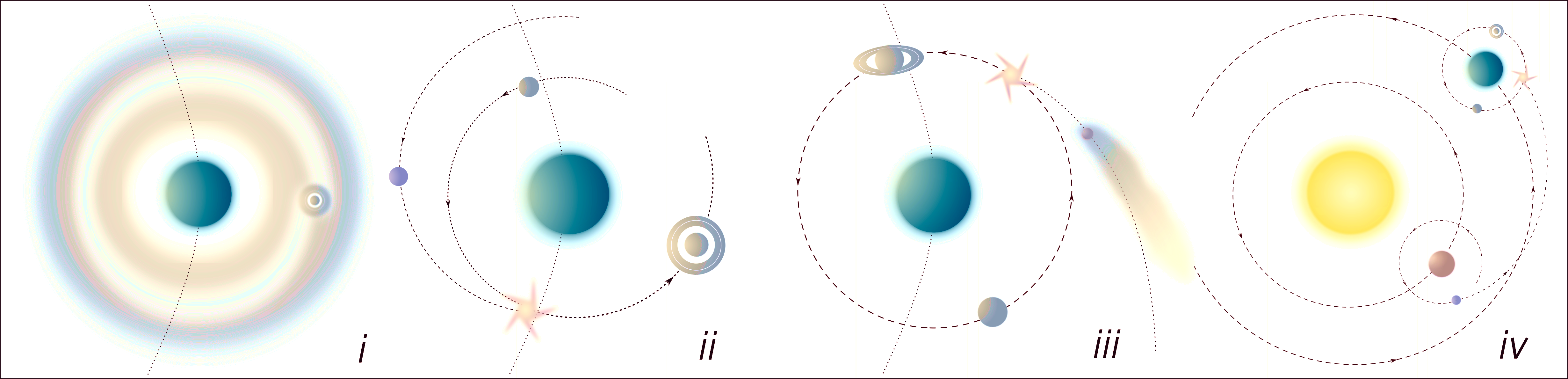}
        \caption{Possible pathways to form \textit{cronomoons}, according to Section \ref{sec:scenarios}: \textit{i)} corresponds to circumsatellital disc remnant, \textit{ii)} to moon-moon collisions, \textit{iii)} to exocomet collisions, and \textit{iv)} to ploonets. Note that the scenarios from \textit{i)} to \textit{iii)} are planetocentric frameworks, whereas \textit{iv)} is centred on the star.}
    \label{fig:cm-form}
\end{figure*}

Using approaches such as pure-dynamical simulations including tidal interactions, the stability of hierarchical systems of moons (i.e. moons around moons) has been proposed and tested (see \citealt{Forgan2018, Kollmeier2019, Rosario-Franco2020}). In this regard, to analyse the stability and lifespan of large sets of particles in disc-like structures around moons, it is important to construct a framework able to explain observations displaying baffling signals. For instance, recent analysis of some light curves have suggested the existence of planets whose densities are abnormally low, also called super-inflated planets (see. e.g. \citealt{Akinsanmi2020, Piro2020}). However, such planets could also have a normal density and be surrounded by systems of rings opaque enough to eclipse a significant amount of the stellar disc (see e.g. \citealt{Zuluaga2015} and references therein). Also, the magnitude and detectability of TTV and TDV signals for different theoretical populations of exomoons have been constrained \citep{Sucerquia2019, Sucerquia2020} and delimited \citep{Kipping2020} in terms of the semi-major axes and masses of systems of moons, indicating that any TTVs or TDVs larger than these constrained values must have a different origin.

Given the unusual characteristics of the exomoon candidate Kepler-1625b i, is it possible that we may be observing a different type of system? May this moon have a ringed structure making it look bigger than it really is? What are the detectable imprints of such a system on the light curve? To answer these questions, we explain in Section \ref{sec:scenarios} some possible scenarios for the formation of hypothetical ringed exomoons, while we test their long-term dynamical and thermodynamic stability in Section \ref{sec:stability}. The detectability of such moons using current and future instrumentation is studied in Section \ref{sec:detectability}, and we analyse in Section \ref{sec:k1625b} the system \rev{Kepler-1625b i} under the scope of this work. Finally, we discuss in Section \ref{sec:discussion} the main findings of this research and its possible consequences on the search and characterisation of exomoons and exorings.

\section{Formation scenarios of \textit{cronomoons} }
\label{sec:scenarios}

Rings seem to be an exclusive feature of the giant planets in the Solar System. Each of them has a set of circumplanetary rings with quite dissimilar characteristics regarding composition, surface density, and optical properties. These differences suggest the existence of different formation and evolutionary pathways of these ring systems (see, e.g. \citealt{Hyodo2017}). Moreover, it is already known that some minor Solar System bodies, such as the centaurs \textit{Chariklo} and \textit{Chiron} as well as the trans-Neptunian object \textit{Haumea}, bear ring system that have been detected by stellar occultations. The formation mechanisms of these rings are still unknown, although it is presumed that they could be the result of a collision between small bodies (see e.g \citealt{Braga-Ribas,Ortiz_2015,Ortiz2017} and citing articles).

The variety of ring species known so far, the fact that small-mass bodies may possess them, plus the stability studies developed in Section \ref{sec:stability} lead us to think that moons may also have a ring system around them. We discuss below on what these formation routes might be and their possible evolutionary paths.

\begin{enumerate}
    \item \textbf{Circumsatellital disc remnant:} 
    One of the main but still debated scenarios to explain the origin of rings-systems around Saturn draws an analogy to circumstellar discs during the process of planetary formation, proposing that Saturn's rings are the remnants of Saturn’s sub-nebula disc \citep*{Pollack1973,Pollack1975}. However, because the average chemical composition of these rings is different from that of Saturn's satellites, it has been argued that the formation of the rings occurred at a later stage, as explained in detail by \citet{Charnoz2009}. 
    
    Still, it is worth noting that growing planets are surrounded by remnant discs which can form regular satellites, and during their formation some parts of the discs can thrive around the satellites (see case \textit{i)} in Fig.~\ref{fig:cm-form}). For instance, a dusty circumplanetary disc with a bulk mass of $\sim10^{-3} \; M_\oplus$ has been recently observed around PDS70 b, and there is evidence of another one surrounding a the planet c, which have been interpreted as a moon-forming region \citep{Isella_2019}. Therefore, planets and regular satellites share an analogous formation channel where discs composed of dust (and sometimes ice) can survive the coalescence of material and last for long time-scales that would allow us to detect them either via their emission properties (as PDS70 b ,c), or the imprints they would produce on the light curve of their host planet/satellite.

    \item \textbf{Moon-moon collisions:}
    The rate at which moons migrate due to tidal angular momentum exchange depends mainly on their masses and the properties of the planet. Supposing a steady state for planetary migration, giant planets harbouring moons with dissimilar compositions and masses (e.g. one dense and rocky, the other light and composed of lighter material) will induce different rates of orbital migration through tidal migration as they are gravitationally perturbed. The worst-case scenario of this combination of gravitational effects could lead the system to chaotic orbital configurations that end in collisions with each other, or between the moons and the planet, leading to the formation of rings of particles around them.
    
    In the Solar System, giant planets have a large number of satellites around them. Similarly, we expect massive exoplanets (at large circumstellar distances) to harbour large numbers of satellites, rings, and exomoons. However, different evolutionary models suggest that short-period \citep{Alvarado2019} and ultra-short-period planets \citep{Alvarado2021} have undergone a slow but progressive migration to reach the currently observed positions. In such processes, different planetary properties will be affected, thereby affecting their surroundings. For example, the planetary Hill radius is compressed \citep*{Spalding2016}, allowing a stronger gravitational interaction between its satellites and other external bodies such as the host star. Also, the angular momentum exchange between planets and their satellites is affected by migration as tidal interaction become stronger, which will modify the planetary rotation rate and the orbital mean motion of moons.
    
    At present, the problem of the orbital stability and fate of multiple moons migrating in conjunction with the planet has not yet been studied in depth, but processes such as those described above could move these moons away from their possible protective resonances and introduce chaos into the system (see e.g. \citealt{Quarles2021}, and references therein), which may result in collisions and mergers between them. The by-product of such interactions could be the depletion of these moons, or the survival of a few of them (or a single massive moon) which transiently possess surrounding ring-like particle systems after the collision (see case {\it ii} in Fig. \ref{fig:cm-form}), and are expected to survive for longer than hundreds of Myrs \citep{Hesselbrock2017,Crida2019}. However, the moon-moon collisions giving rise to such rings are stochastic processes that are not yet constrained by observations, with a lack of stringent theoretical computations from population synthesis studies. Whether the rings resulting from these collisions are made of dust or ice depends on the composition, density and their reservoirs of icy material, which will be given by the distance of the host planet from the star.
    
    \item \textbf{Exo-cometary collisions:}
    Currently, cometary activity has been detected in about thirty stars. These objects have been identified, on the one hand, due to their asymmetric light curves that reveal the presence of their tails, and on the other hand, because of the chemical fluctuations from starlight passing through their tails and transiently modifying the stellar spectrum (for a complete summary of these systems see \citealt{Strom_2020}).
    
    In the Solar System, long-period comets emerged after the orbital exchange of Neptune and Uranus described in the Nice model \citep{Tsiganis2005}. Exoplanetary systems are also prone to processes where systems of low-mass icy or dusty objects have high-eccentricity orbits around their star. Under the presence of short-period planets retaining their moons \citep{Namouni2010, Spalding2016}, it is likely that exocomets will collide with such planets and moons (this scenario will be the subject of a future work, led by the same authors of this paper). Also, after a long-lasting gravitational interaction with the planet, such exocomets may end up located in short-period orbits which increase the probability of collision to form tilted rings around planets or moons. Icy exocomets colliding against moons and planets motivated the calculations done in this work using icy particles as possible constituents of close-in exorings (see case {\it iii} in Fig.~\ref{fig:cm-form}), whose life time-scales will be computed in Section \ref{sec:stability_therm}. 

    \item \textbf{Ploonets:} 
    Ploonets are moons that have been moved from circumplanetary to circumstellar orbits by the effect of, for example, tides (\citealt{Sucerquia2019}). These ploonets manage to settle transiently into orbits close to their host planet, but they may have eccentric orbits. These orbital characteristics make them very prone to collide either with their host planet or other nearby planets and moons,  as depicted in the middle panel of Fig.~\ref{fig:cm-form}, which is a scenario that we will study in a future paper. Depending on the orbital inclinations of ploonets, collisions and close approaches could form rings-systems around planets and their moons.
    \item \rev{ \textbf{Other formation channels:} 1. A common feature between the giant planets of the Solar System is the presence of irregular satellites whose origin is explained via gravitational capture of wanderer objects in the planetary neighbourhood. If rings around bodies such as Chariklo, Chiron, and Haumea survive to the capture process of giant planets, this might form a \textit{cronomoon} in the Solar System. 2. Under the assumption of the existence of \textit{submoons} \citep{Kollmeier2019}, in analogy to satellital migration around short-period planets, a submoon could undergo inward migration until reaching its Roche limit with respect to the host moon, spreading its material and creating a system of rings \citep{Canup2010}.}
\end{enumerate}

\section{Ring's lifespan, shape, and size}
\label{sec:stability}

%FFFFFFFFFFFFFFFFFFFFFFFFFFFFFFFFFF
%FFFFFFFFFFFFFFFFFFFFFFFFFFFFFFFFFF
\begin{figure}
	\includegraphics[width=\columnwidth]{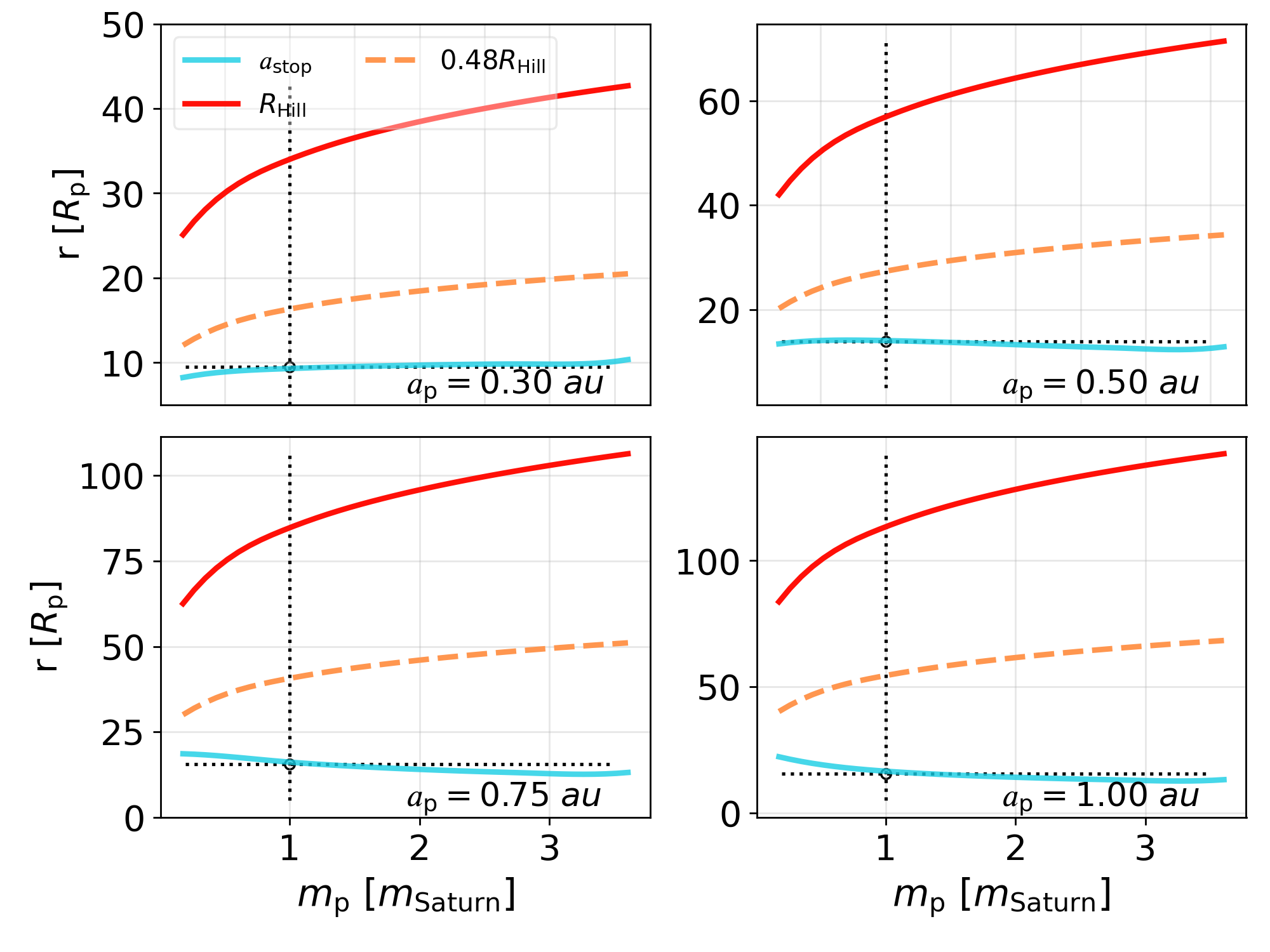}
	\includegraphics[width=\columnwidth]{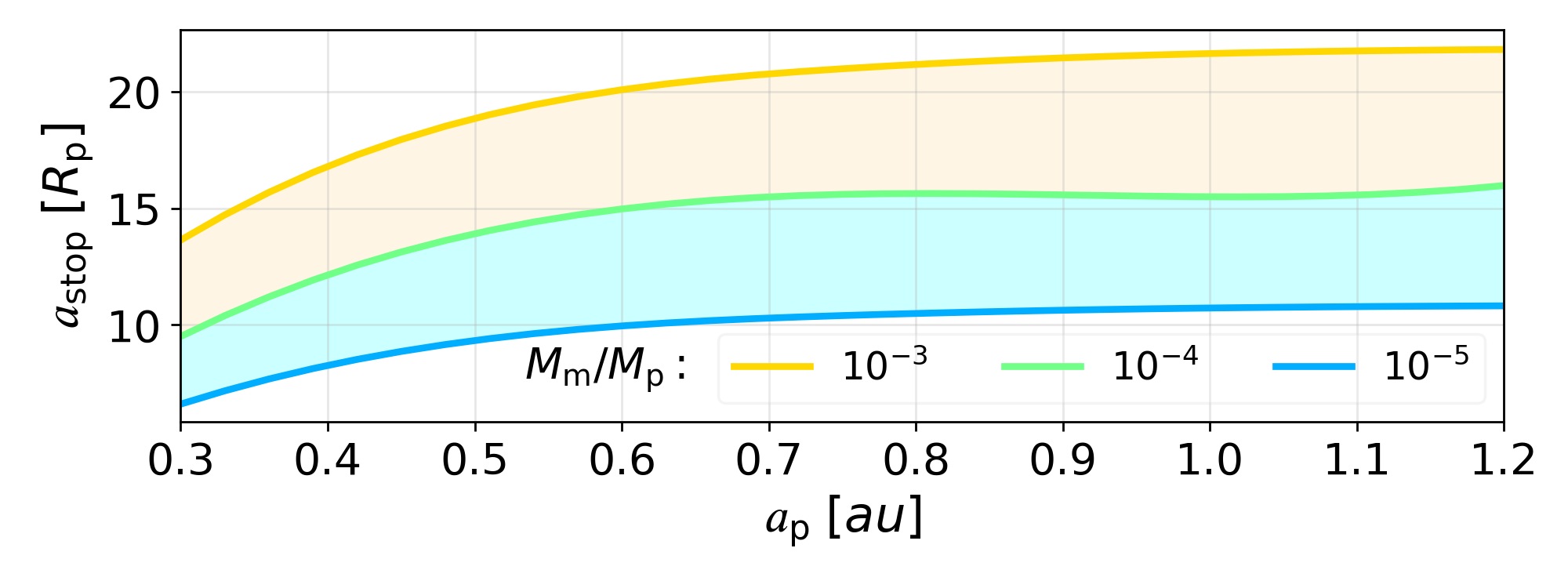}
    \caption{Upper panel: asymptotic moon's semi-major axis ($\astop$) as a function of the planetary mass (blue solid lines). For reference, the principal and secondary (stability limit) Hill radii are drawn in red and orange lines, respectively. The intersection between dotted lines are the values used in our simulations (Section \ref{sec:stability_dynam}). Lower panel: final position $\astop$ for planet--moon systems having different moon-to-planet mass ratios. We assume here the physical properties of a Jupiter-like planet and a Titan-like moon.}
    \label{fig:constraints-am}
\end{figure}
%FFFFFFFFFFFFFFFFFFFFFFFFFFFFFFFFFF
%FFFFFFFFFFFFFFFFFFFFFFFFFFFFFFFFFF

In this section, we test the dynamical stability of a system composed of a \rev{Solar-like host star} and a close-in giant planet orbited by a massive ringed moon (see Fig.~\ref{fig:scheme} and the nomenclature therein used throughout this work). Planet-moon mass ratios are assumed to be within the framework of Galilean's moons formation, where $\Mmoon/\Mp \sim 10^{-4}$. Also, moon's orbital positions (i.e, moon's semi-major axes) will be constrained by adopting the satellite migration model around close-in exoplanets described by \cite*{Alvarado2017} and tested in \cite{Sucerquia2020} (see Section \ref{sec:tidal}), allowing us to reduce the number of unknown parameters. 

\begin{figure*}
	\includegraphics[width=\columnwidth]{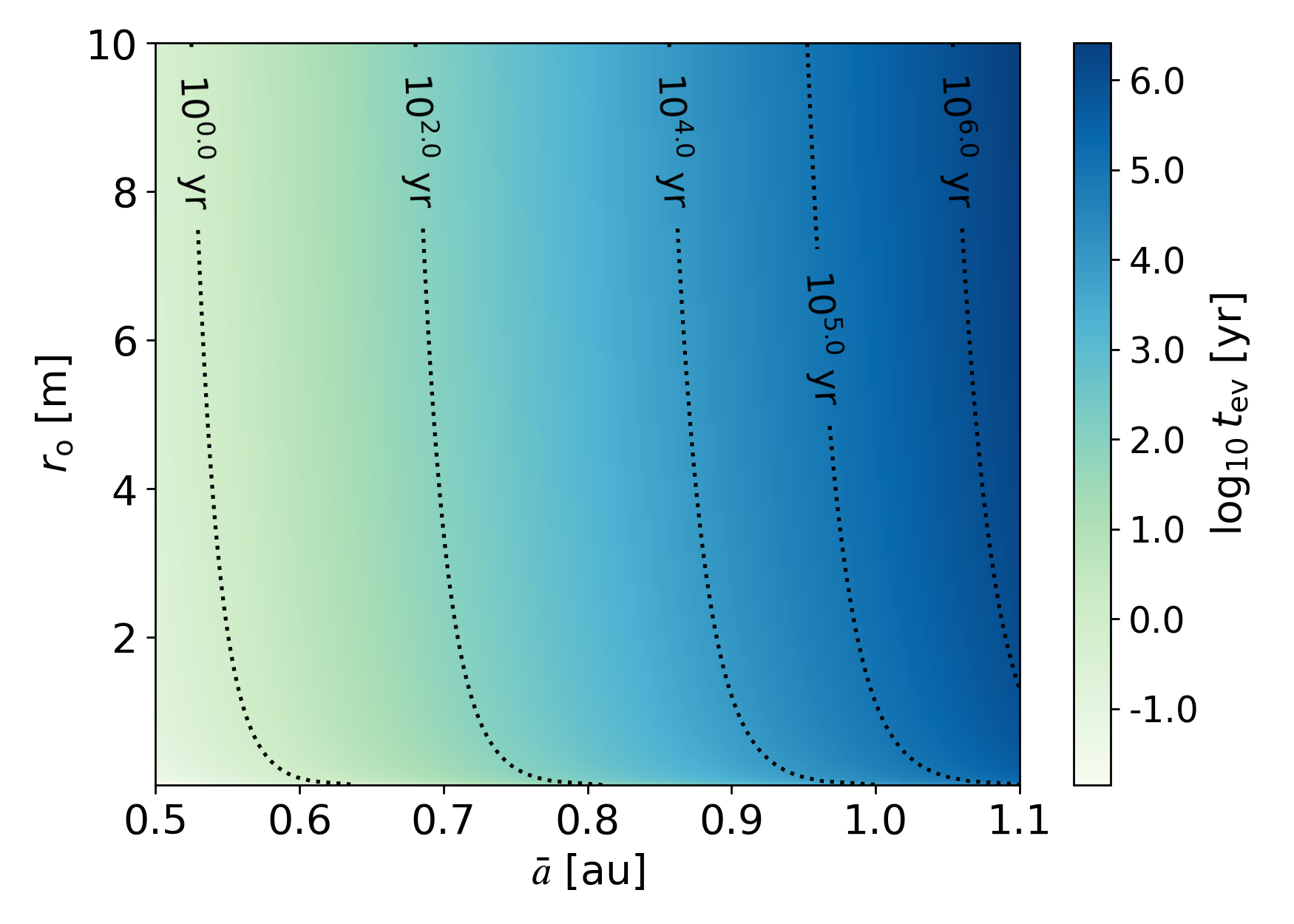}
	\includegraphics[width=\columnwidth]{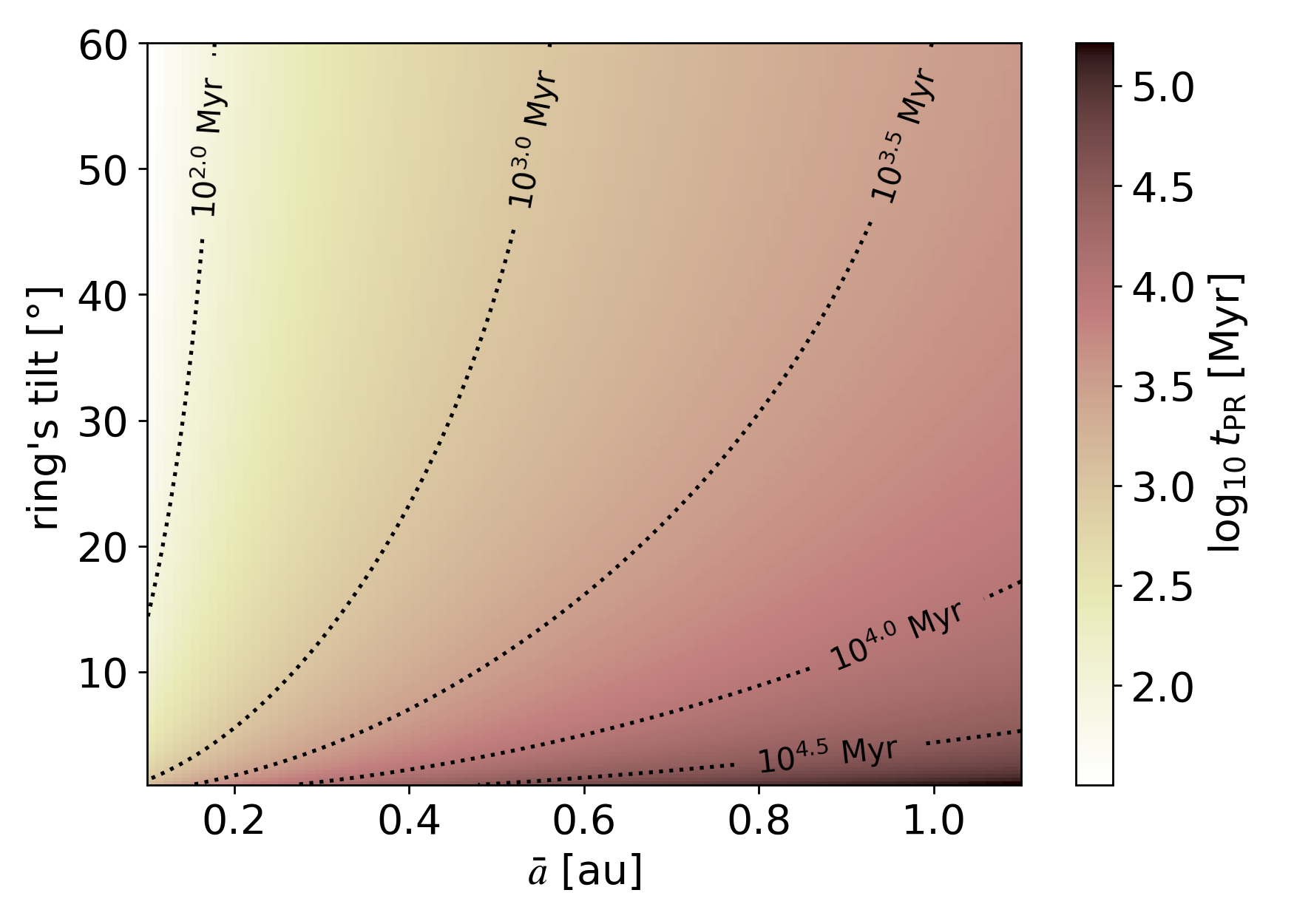}
    \caption{Left-hand panel: Sublimation time-scales for water--ice spheres (with bond albedos of $0.95$) of different sizes and placed at average distances $\bar{a}$ from the star. Right-hand panel: Poynting-Robertson decay time-scales for different tilts of the rings, placed at $\bar{a}$ from the star.}
    \label{fig:timescale-pr}
\end{figure*}

Seeking a more significant contribution of \textit{cronomoons} in planetary light curves, we set up our simulations so that planets have close-in circular orbits from their host star, with distances ranging from 0.3 to 1.0 au. At such orbital positions the system is exposed to large amounts of stellar radiation, which makes volatile material highly unstable. For reasons explained in Sections \ref{sec:stability_therm}, \ref{sec:detectability}, and \ref{sec:discussion}; and in spite of the high irradiation received by the system, we will not restrict our study to rings composed of refractory material (the so called \textit{warm-exorings} described by \citealt{Schlichting2011}); that is, we will also include those composed of ices. In this work we have dubbed them dusty and icy rings when composed of silicates and ices, respectively.
%Because the vicinity of the system to the star we have assumed that ring's particles are made of thick refractory material instead volatile one, becoming so a subset of the hypothetical system of the so called \textit{warm-exorings} described by \citealt{Schlichting2011}. 

\subsection{Tidal migration of host moons}
\label{sec:tidal}

To study the dynamical stability of \textit{cronomoons} we need to constrain the orbital and physical features of host moons. The semi-major axis of the moon will be assumed at a position also known as Satellite's Tidal Orbital Parking ($\astop$, \citealt{Sucerquia2020}), reached due to orbital migration triggered by tidal interactions with the parent planet. According to the model in \citet{Alvarado2017}, which we adopt in this work prior to the evolution of \textit{cronomoons}, both the evolution of physical size (i.e. the planetary radius $\Rp$) and interior dissipative properties of the planet modify the final fate of migrating moons. This can have important consequences for the survival of exomoons in compact systems and even for the orbital evolution of short-period \citep{Alvarado2019} and ultra-short-period planets \citep{Alvarado2021}.

For this work, as some important criteria for orbital migration of moons from \citet{Alvarado2017} share some of its attributes with previous classical models (e.g. \citealt{Barnes2002} and \citealt*{Sasaki2012}), the evolution of the planetary rotational rate ($\Op$) and the planet and moon's mean motion ($\npp$ and $\nmm$, respectively) follow:

\beq{eq:dOdt}
\begin{split}
\frac{\Der\Op}{\Der t}=-\frac{3}{2}\frac{\Kpp}{\Qp}\frac{\Rp^{3}}{\kappa^{2}G}
\left[
\frac{(G\Mstar)^{2}}{\Mp\apos^{6}}\,\sgn(\Op-\npp)+\right.\\
\left.\frac{\Mmoon^{2}}{\Mp^{3}}\nmm^{4}\,\sgn(\Op-\nmm) \right].
\end{split}
\eeq
%\vspace{-1.0cm}
\beq{eq:dndt}
\frac{\Der \nmm}{\Der t}=-\frac{9}{2}\frac{\Kpp}{\Qp}\frac{\Mmoon\Rp^{5}}{ G^{5/3}\Mp^{8/3}}\nmm^{16/3}\sgn(\Op-\nmm).
\eeq
Where the moon's semi-major axis ($\amoon$) can be found via Kepler's third law as $\amoon\approx(G\Mp/\nmm^2)^{1/3}$ and $\kappa$ is the planetary gyration radius. Also, the dissipative properties of the planet are represented by the Love number $\Kpp$, a coefficient standing for its deformation capabilities; and the tidal quality factor $\Qp$, which represents the tidal energy dissipated per rotational period. \rev{For both $\Kpp$ and $\Qp$ we use the formalism of \citet{Alvarado2017} which follows \citet{Ogilvie2013} and \citet{Guenel2014}. Within this formalism we assume initial $\Kpp$ and $\Qp$ as those of a Jupiter-like planet \citep{Mathis2015a} and make both quantities evolve during the orbital migration of moons. This locates moons at a final stable semi-major axis \citep{Sucerquia2020}. On the contrary, if both of these quantities are assumed static, moons would migrate inwards and eventually cross their Roche limit as in \citet{Barnes2002} and \citet{Sasaki2012}}.

Tidal models state that moons can migrate outwards until they reach an orbital position where synchronisation occurs, i.e. $\Op\approx\nmm$. Then, inward migration follows since the torques on the tidal bulge of the planet are reverted. However, in \citet{Alvarado2017} inward migration is suppressed because $\Op$ never reaches $\nmm$, and this might have notable consequences for the moon's final fate, including but not limited to: detachment from the host planet \citep{Sucerquia2019}, or orbital parking on an asymptotic semi-major axis \citep{Sucerquia2020}. We calculate $\astop$ under the model by \citet{Alvarado2017}, and assume a departing moon's semi-major axis of $\amoon = 3\,\Rp$ and a planetary rotational period of 13 h. $\astop$ is the value we use to set up the numerical simulations (Section  \ref{sec:stability}) and to assess their detectability (Section \ref{sec:detectability}). The main outcomes of this section are presented in Fig.  \ref{fig:constraints-am}: the upper panel presents the evolution of $\astop$, the planetary Hill radius $\Rh$, and the stability limit of $0.48\,\Rh$ \citep*{Domingos2006} for planets at $0.3$, $0.5$, $0.8$ and $1.0$ au; while the lower panel shows the relationship between $\ap$ and $\amoon$ for $\Mmoon/\Mp = 10^{-5}, 10^{-4}$ and $10^{-3}$, from blue to yellow, respectively.

\subsection{Thermodynamic stability: rings' lifespan}
\label{sec:stability_therm}

In general, isolated planetary rings are gravitationally stable and their spatial domain is externally constrained by the Roche Limit\footnote{This is defined as the orbital limit inside which a body's self-attraction (particles) is exceeded by the tidal forces of a primary body (host moon).}, $\aroche$, to prevent any rings' particles to form larger bodies \citep{Roche1849, Crida2012},
\begin{equation}
    \aroche = r_\mathrm{m} \left( 2 \frac{\rho_\mathrm{m}}{\rho_\mathrm{part}}\right)^{\frac{1}{3}}
    \label{eq:roche}
\end{equation}
where $r_\mathrm{m}$ and $ \rho_\mathrm{m}$ are the radius and density of the satellite, respectively; and $\rho_\mathrm{part}$ is the density of the ring's particles. Still, at short distances from the star, it is expected that icy rings are promptly sublimated leaving behind rings composed by dusty particles (`dusty rings'). Such lifespan will be limited by the Poynting-Robertson effect. Below, we present the framework to calculate both effects in the context of this work.

\subsubsection{Sublimation of icy rings}
\label{sec:stability_thermo}

%################ FROM PLOONETS  ################

Given that rings around moons can be composed of particles with icy envelopes, we study the evolution of spherical blocks of ice adopting the model proposed in \citet{Sucerquia2019}. This will allow us to estimate the typical time-scales of icy rings' orbital evolution around close-in giant planets, while calculating the sublimation of icy envelopes to constrain the typical lifespan of this type of rings systems. We assume these particles to be rapidly rotating spheres (rotation period much shorter than orbital period) located in circular orbits. The equilibrium temperature of these particles is given by:
\begin{equation} 
\Teq=\left[\frac{\Lstar \, (1-A)}{16\,\pi\,\SigmaSB \,a^{2}}\right]^{1/4},
\end{equation}
where $A$ is the bond albedo, $\Lstar$ the stellar luminosity, $a$ the orbital radius, and $\SigmaSB$ the Stefan–Boltzmann constant. In addition, water from the icy particles sublimates at temperature $T$ at a rate $Z_\mathrm{r}$ (in kg m$^{-2}$ s$^{-1}$) given by \citep{Estermann1955,Bohren1998},

\begin{figure*}
     \centering
        \includegraphics[scale=0.34]{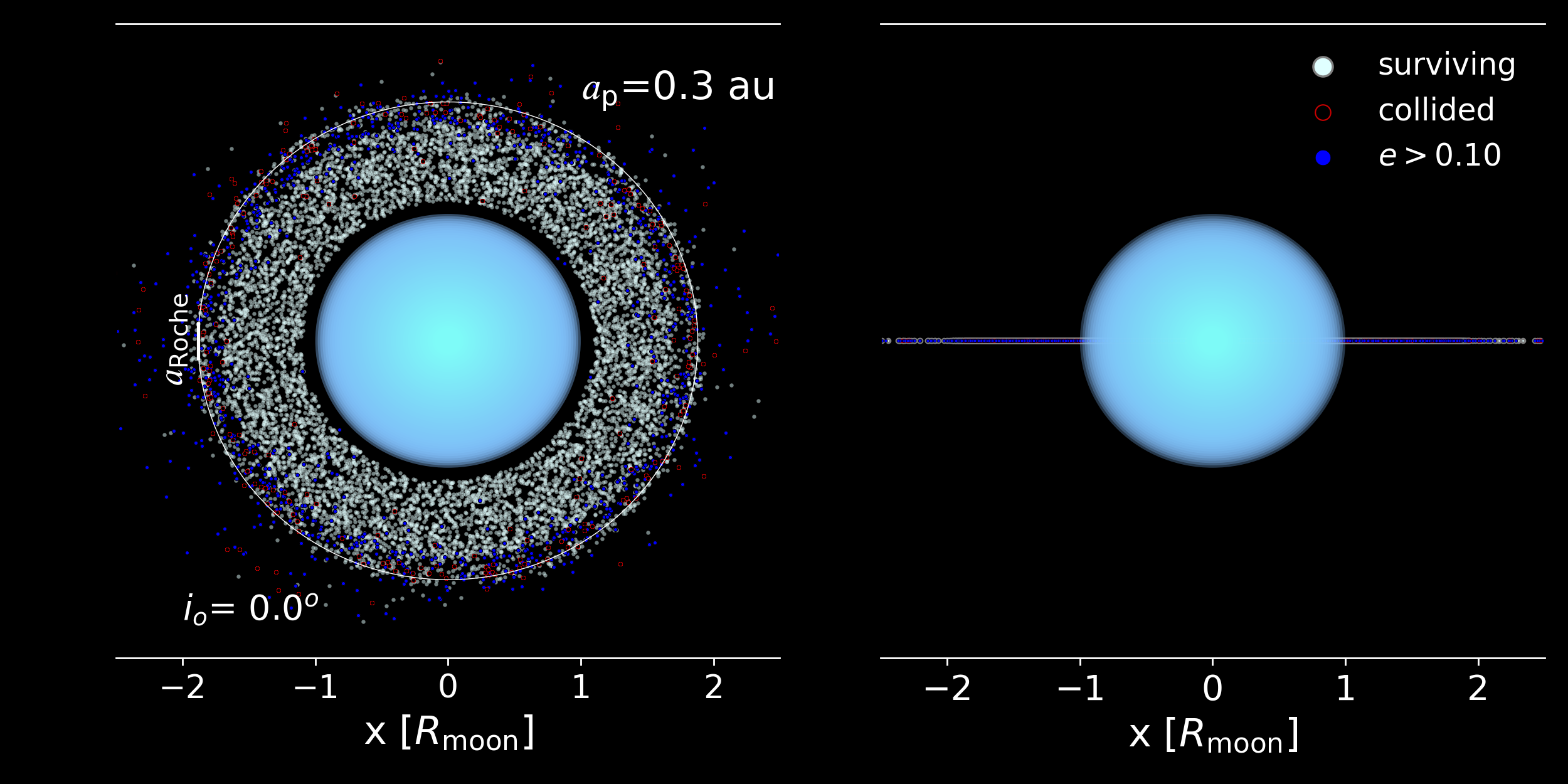}
        \includegraphics[scale=0.34]{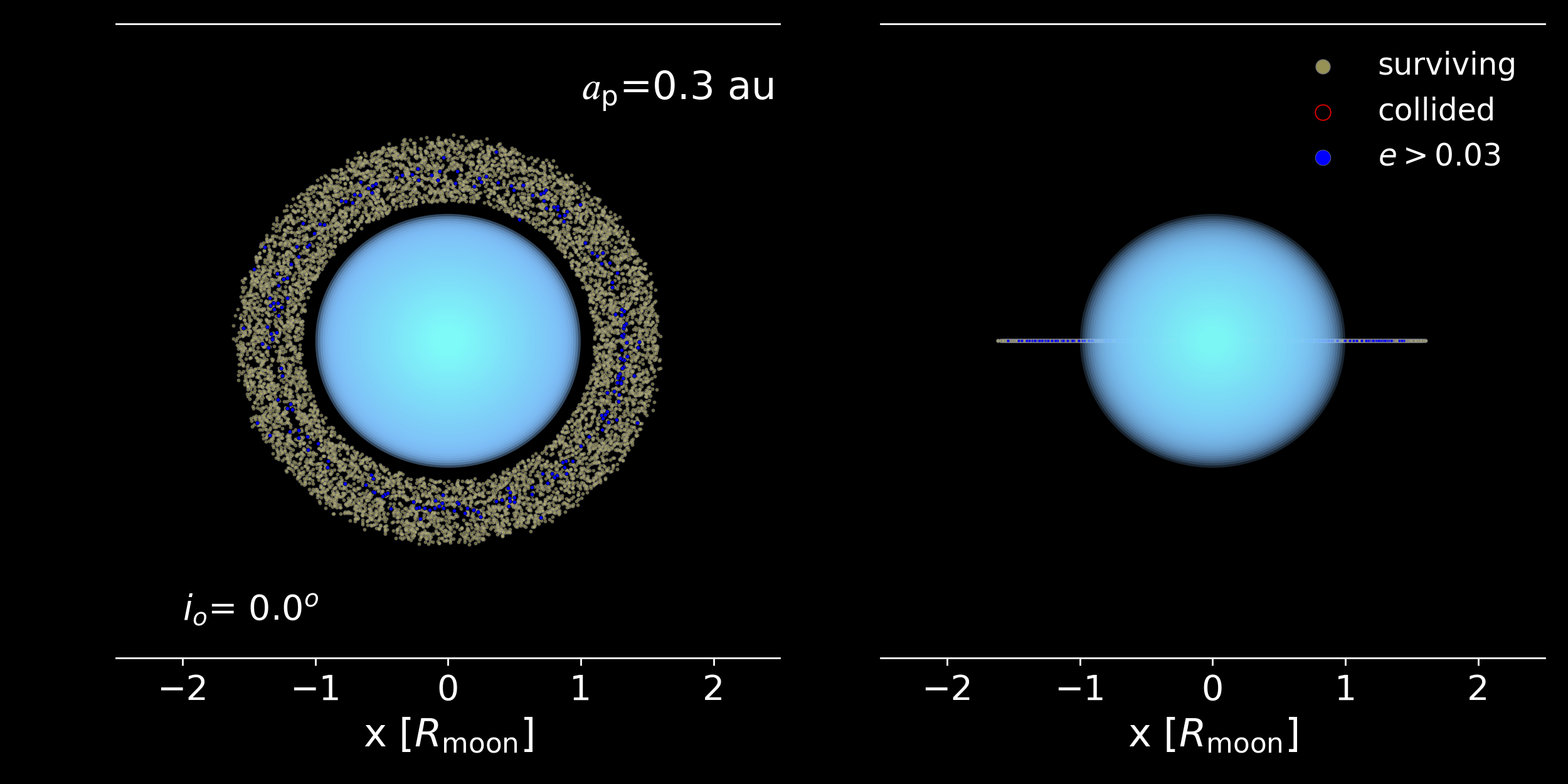}\\
        \includegraphics[scale=0.34]{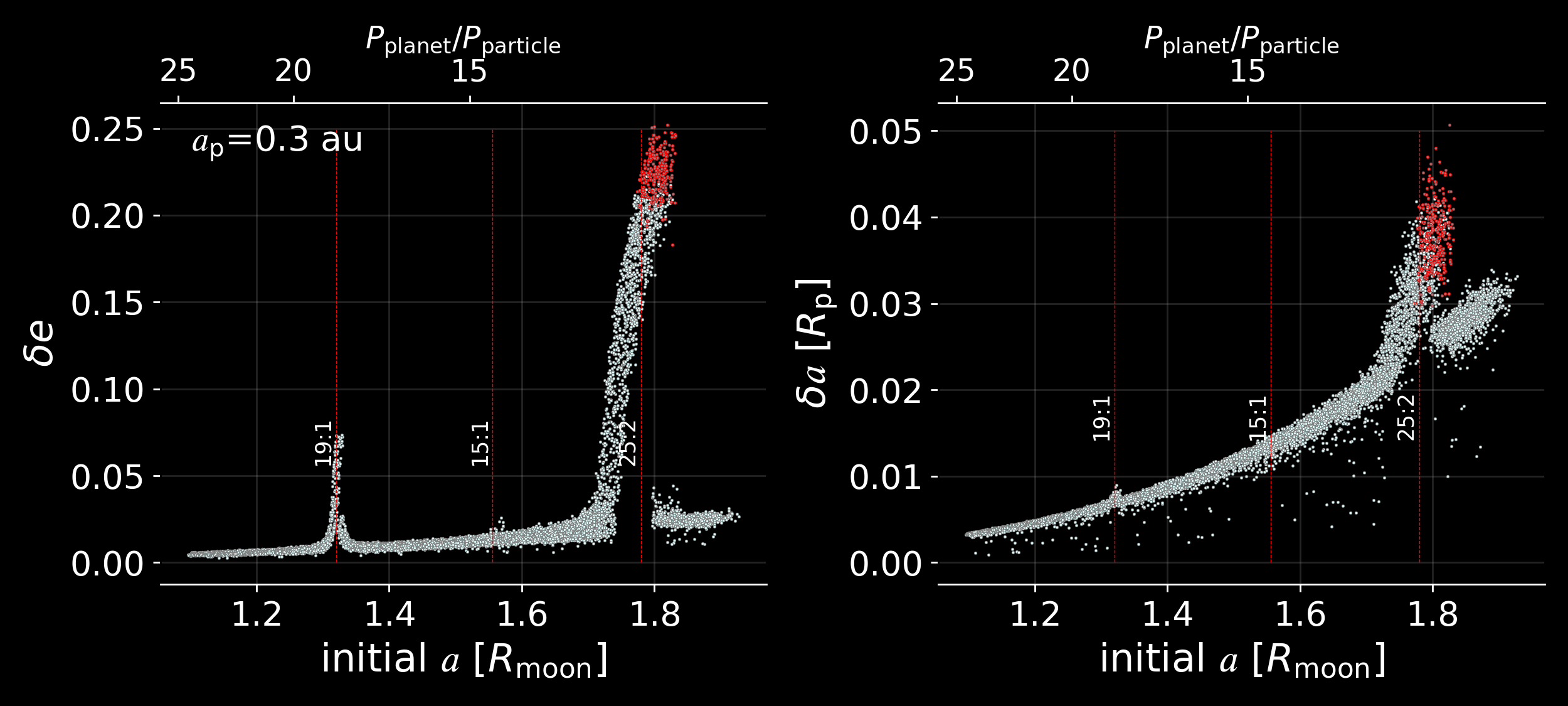}
        \includegraphics[scale=0.34]{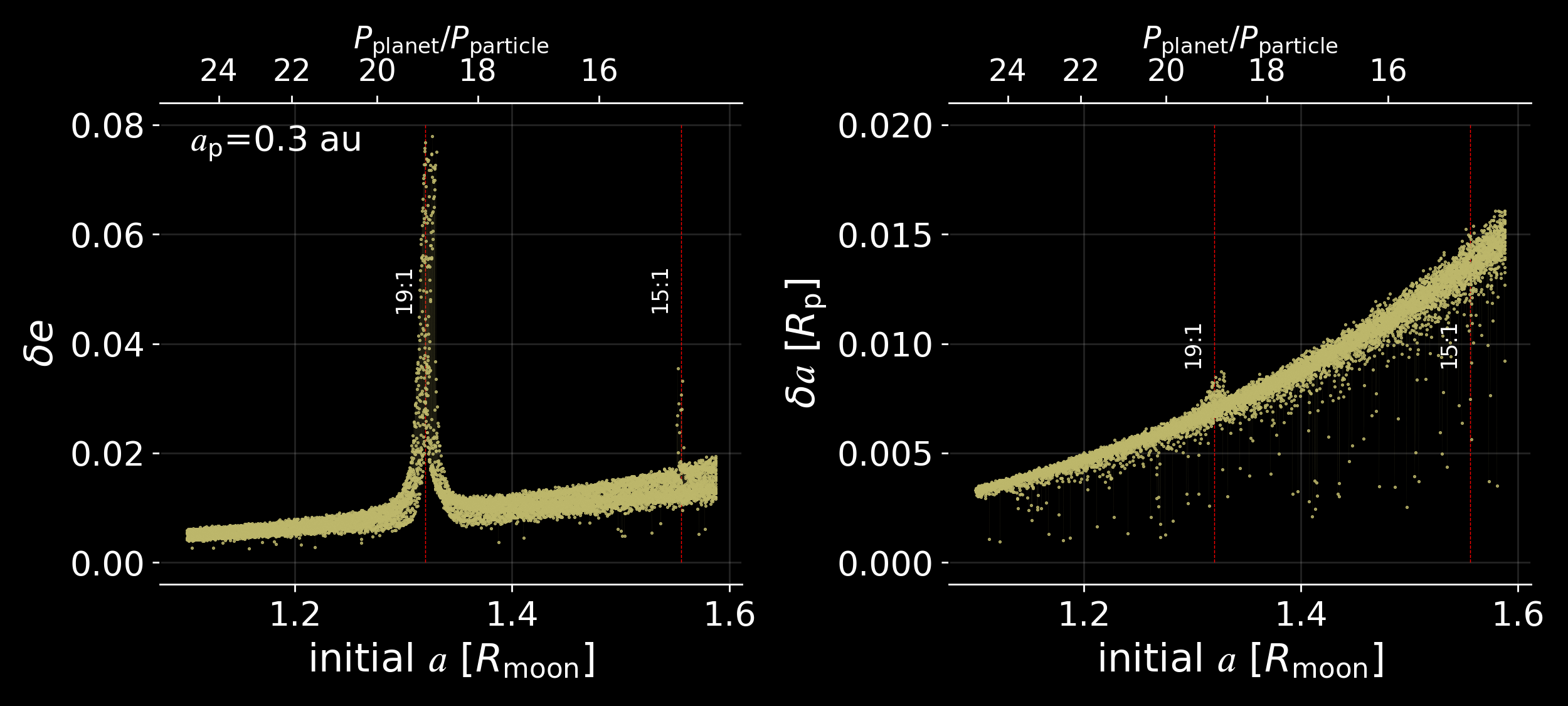}
        \caption{Comparison between icy (left-hand panels) and dusty (right-hand panels) rings for a moon orbiting a planet following a circular orbit at $\ap = 0.3$ au. The rings have an initial inclination of $i=0$° and circular orbits (i.e. $e=0$). The lower panels show $\delta e$ and $\delta a$ of the rings' particles, which are the maximum semi-amplitude of their eccentricity and the semi-major axis, respectively. The state showing the surviving (white), collided (red), and eccentric (blue) particles correspond to a system evolution of 100 yr.}
    \label{fig:comparison}
\end{figure*}

\begin{equation}
\label{eq:z}
Z_\mathrm{r} = e_\mathrm{sat}(T)\left(\frac{m_\mathrm{w}}{2\pi RT}\right)^{1/2}\exp\left[\frac{2\, m_{w}\,\sigma_{i}\,(T)}{\rho_{i}(T)\, r \,RT}\right] \,\,\,.
\end{equation}
with $e_\mathrm{sat}(T)$, $\sigma_\mathrm{i}(T)$, $\rho_\mathrm{i}(T)$ the saturation vapour pressure, the surface ice tension, and  the surface ice density, respectively. $m_\mathrm{w}$ is the molecular weight of water and $R$ the universal constant of ideal gas. All these quantities are adopted from standard thermodynamic expressions by \citet{Andreas2007}, which have widely been applied to study sublimation processes in bodies like the Moon, Saturn's rings \citep{Hedman2015}, and comets \citep{Vincent2016}. With this in mind, the total water mass loss rate, $\dot{M}_\mathrm{s}$, can be calculated as follows:
\begin{equation} 
\label{eq:Msurf-rate}
\dot{M}_\mathrm{s} = - Z_\mathrm{r}S\sub{pl}(T) = -Z_\mathrm{r}\left[ \frac{6\,\sqrt \pi\,m}{\rho_\mathrm{i}(T)}\right]^{2/3},
\end{equation}
where the sphere's surface area $S\sub{pl}(T)$ has been written in terms of its mass $m$ and $\rho\sub{i}(T)$ (in kg m$^{-3}$), which is a function of temperature:
\begin{equation}
    \rho\sub{i}(T) = 916.7 - 0.175\,(T - 273.15) - 5.0 \times 10^{-4} (T-273.15)^2 
\end{equation}
and agreeing within about 1\% of the tabulated values for ice density for temperatures from 13 to 273.15 K \citep{Hobbs1974}. Then, the temporal evolution of the ice block's mass reads:
\begin{equation}
m^{1/3}-m\sub{o}^{1/3} = Z_\mathrm{r} \left( \frac{1}{\rho\sub{i}(T)}\right)^{2/3} \left(\frac{4\pi}{3} \right)^{1/3} \; (t\sub{o}-t), 
\end{equation}
where the initial mass  $m\sub{o}$ corresponds to an initial time $t\sub{o}$. Thus, the time necessary to fully sublimate the particles' ice envelope is:
\begin{equation}
\label{eq:tev}
    t\sub{ev} =  \frac{1}{Z\sub{r}}\left( \frac{4\pi \; m\sub{o}}{3\;\rho\sub{i}(T)^2} \right)^{1/3}.
\end{equation}

The lifespan of icy rings around exomoons is presented in the left-hand panel of Fig.~\ref{fig:timescale-pr}. There, the constituent particles of the ring are assumed to be spheres with radii ranging from $1$ to $10$ m (parallel to the particles in Saturn's rings, see \citealt{Brilliantov9536}) and placed at semi-major axes $\bar{a}= 0.5-1.0$ au. Note that distances $ \bar {a} <0.6 $ au are harmful to icy rings, with corresponding lifespans of less than a year. \rev{However, less luminous stars might allow {\it cronomoons} with icy rings and orbiting close-in planets to exist for longer time-scales}. Also, icy rings located at larger distances can survive up to time-scales of the order of Myr. In fact, the calculations presented in Fig.~\ref{fig:timescale-pr} correspond to an underestimation of the actual values for ice sublimation (i.e. we are not taking into account the actual shape of the particles): irregular or asymmetric surfaces would produce regions where the incidence and absorption of stellar radiation is not homogeneous, affecting thus the sublimation of the rings' particles. Furthermore, when rings' tilts are either low or co-planar to the moon's orbit, the shadow projected on the rings' particles from the planet and the moon will shield the rings from stellar radiation.
\vspace{-0.5cm}

\subsubsection{Poynting-Robertson decay}

We assume that once icy rings are depleted of volatile material, only dusty particles are left in the ring. Then, the Poynting-Robertson effect acts on these dusty particles and triggers their orbital decay and lost. The time-scales for the Poynting-Robertson particle decay ($\tpr$) for `warm-exorings', as derived by \cite{Schlichting2011}, reads:
\begin{equation}
\tpr \sim \frac{\pi c^2 \Sigma}{\sin{i} \: (\Lstar/4\pi \ap^2)\; \Qpr(5+\cos{i}^2)}
\label{eq:therm}
\end{equation}
where $\Sigma= 109$ g cm$^{-2}$ is the mass surface density of the ring, assumed similar to Saturn's B-ring \citep{HEDMAN2016}, $c$ the speed of light, $i$ the ring's tilt (i.e. normal angle with respect to the planet's orbital plane), $\ap$ the planetary semi-major axis, and $\Qpr$ the radiation pressure efficiency factor which was assumed equal to 0.5. The right-hand panel of Fig.~\ref{fig:timescale-pr} shows the value of $\tpr$ for several combinations of averaged star--moon distances $\bar{a}$ and ring's tilt $i$. According to Equation \ref{eq:therm} and both panels of Fig.~\ref{fig:timescale-pr}, ringed moons with high inclinations can have lifespans of the order of Myr.

\label{sec:stability_dynam}

\begin{figure*}
     \centering
          \begin{subfigure}[b]{\textwidth}
        \centering
        	\hspace{0.15cm}
        \includegraphics[scale=0.33]{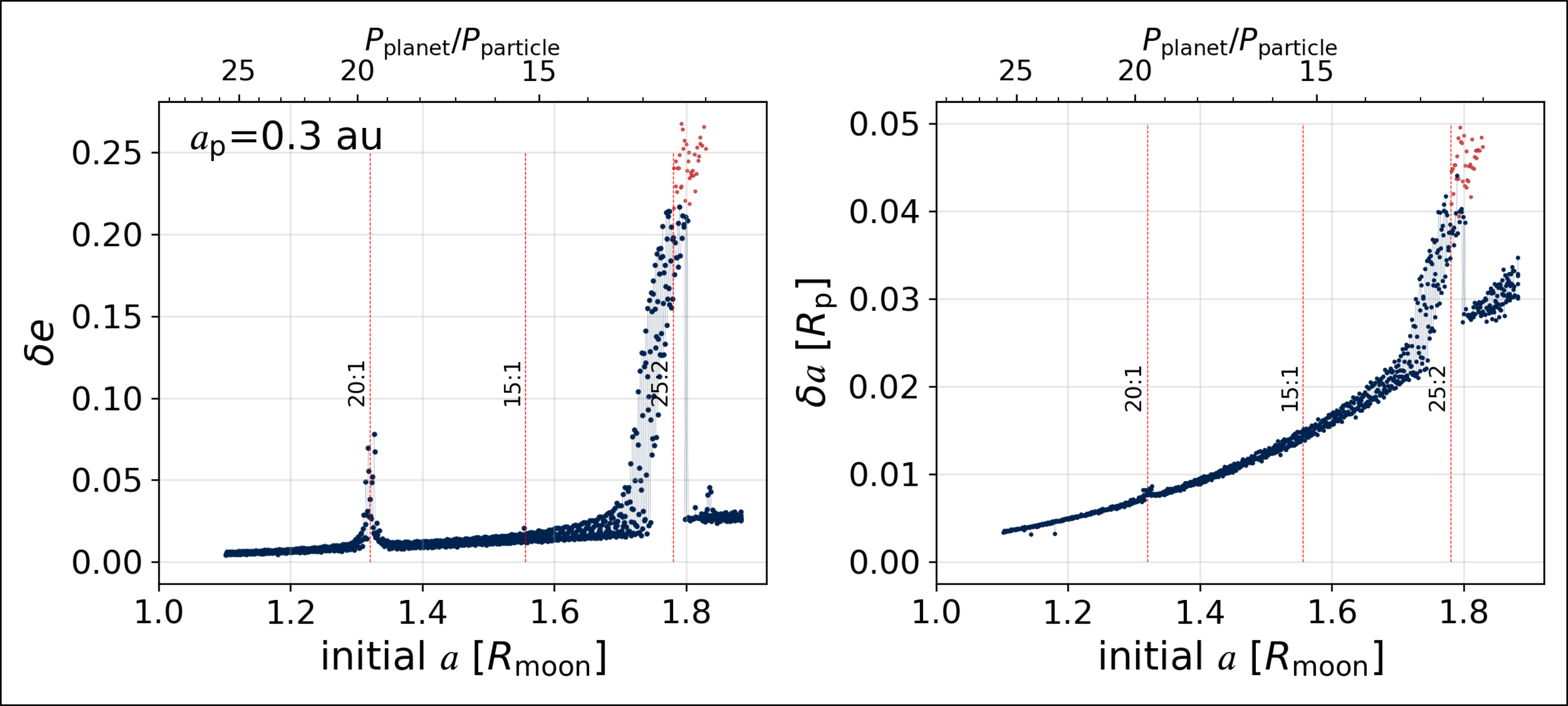}
            \hspace{0.15cm}
        	\vspace{0.25cm}
        \includegraphics[scale=0.33]{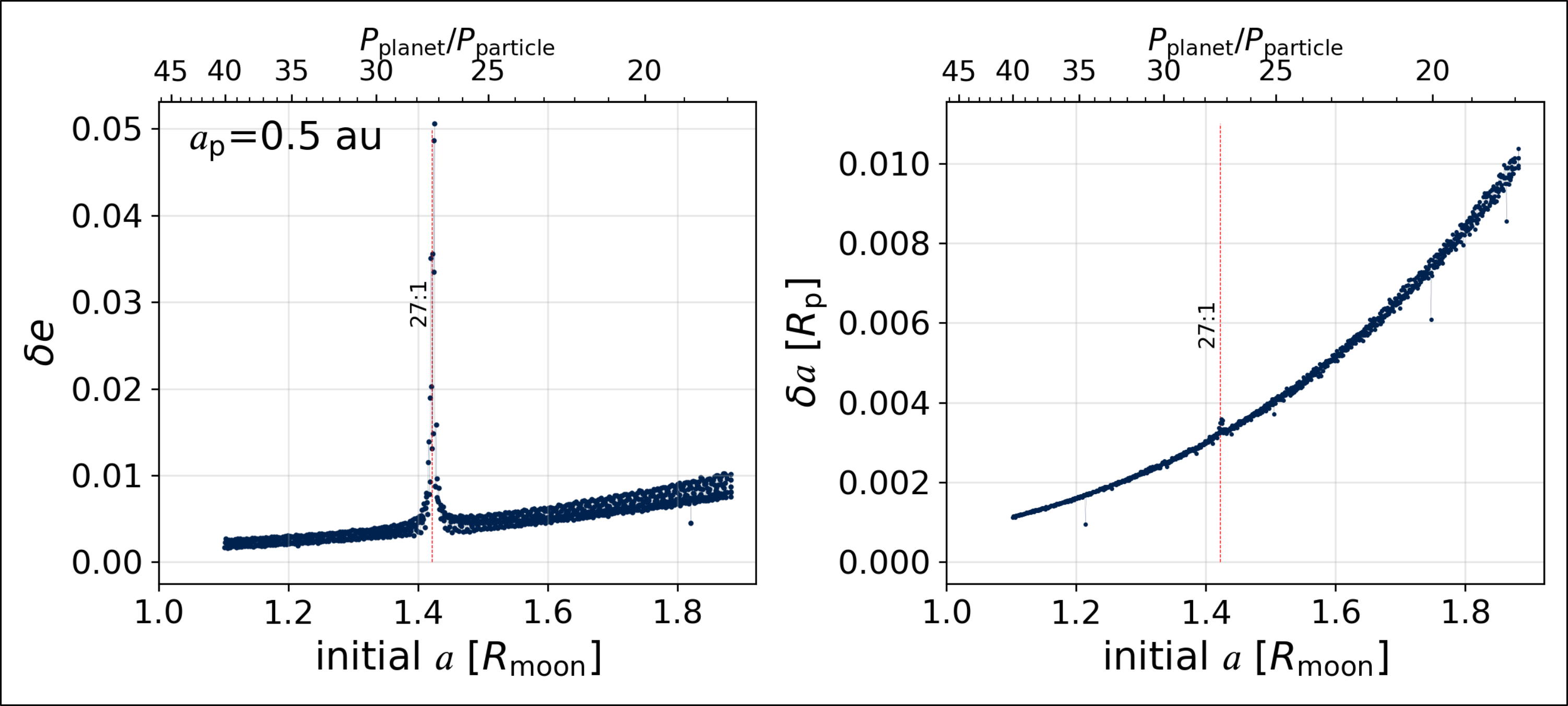}
     \end{subfigure}
          \hspace{0.15cm}
     \begin{subfigure}[b]{\textwidth}
        \centering
        	\hspace{0.15cm}
        \includegraphics[scale=0.33]{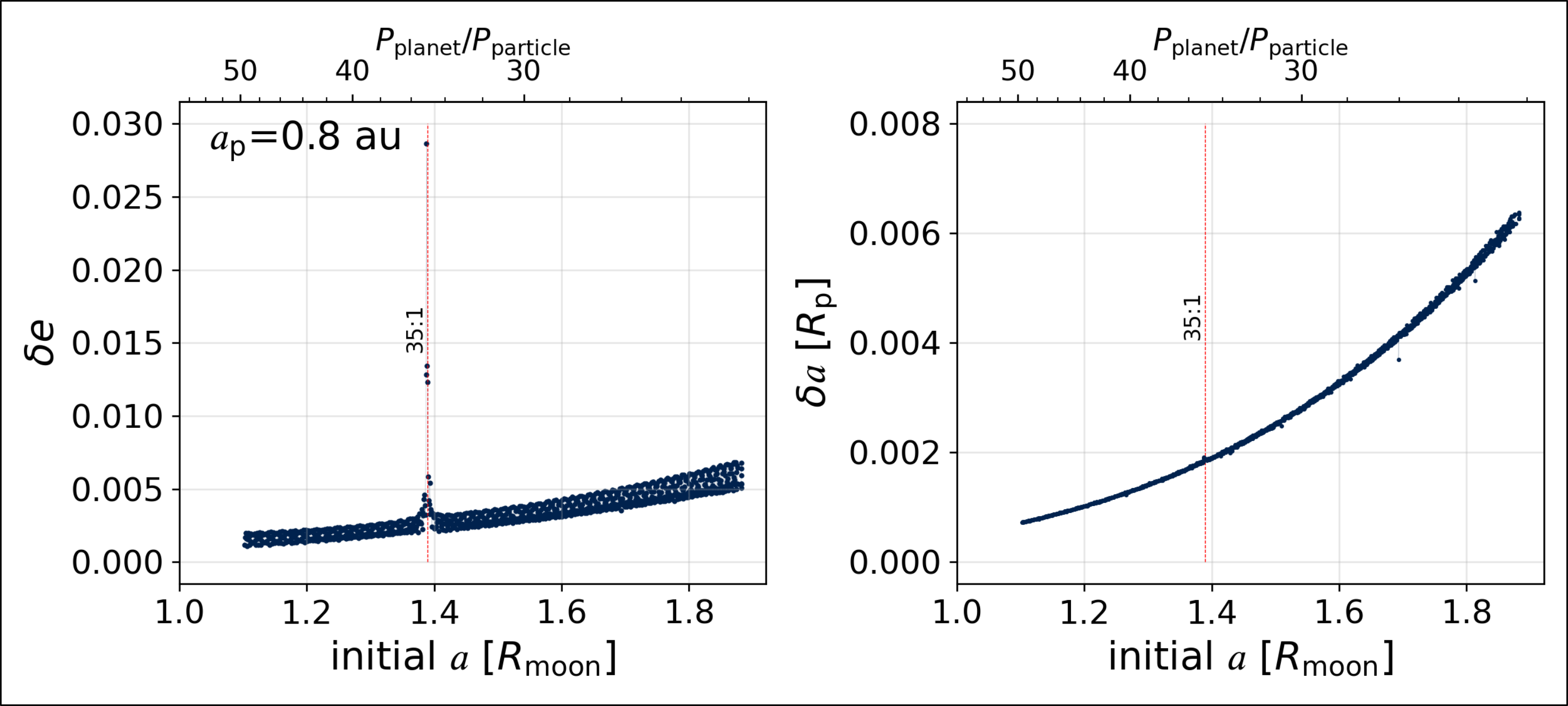}
        	\hspace{0.15cm}
        \includegraphics[scale=0.33]{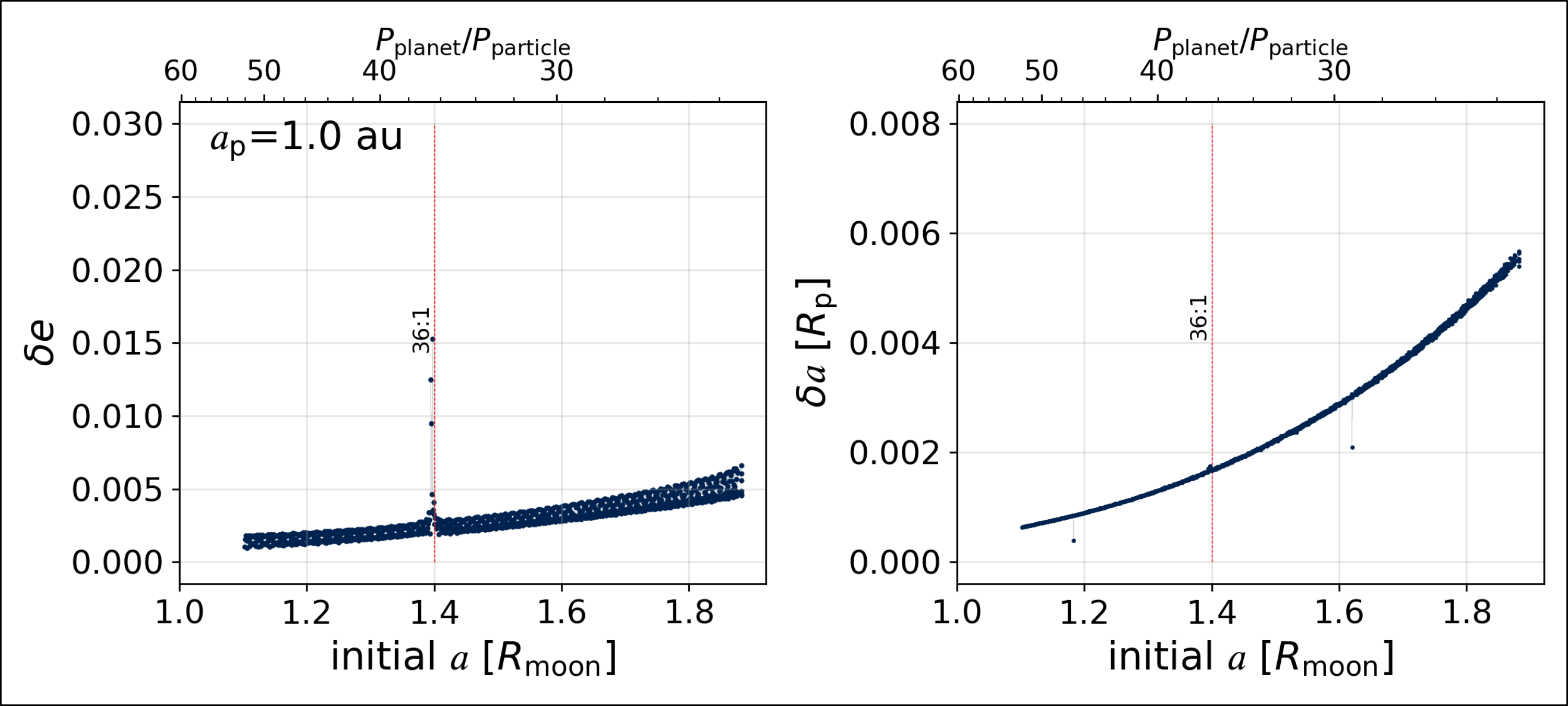}
     \end{subfigure}
     \hfill
     \caption{Numerical simulation outcomes after $\sim 750.000$ periods (in average) for icy rings having a tilt of $0^\circ$. $\delta a $ and $\delta e$ is the maximum semi-amplitude of the excitation of the ring particles' semi-major axis and eccentricity, respectively. Points in red are those particles that collided with the satellite.}
    \label{fig:ringsi0}
\end{figure*}

\begin{figure*}
     \centering
          \begin{subfigure}[b]{\textwidth}
        \centering
            \hspace{0.15cm}
        \includegraphics[scale=0.33]{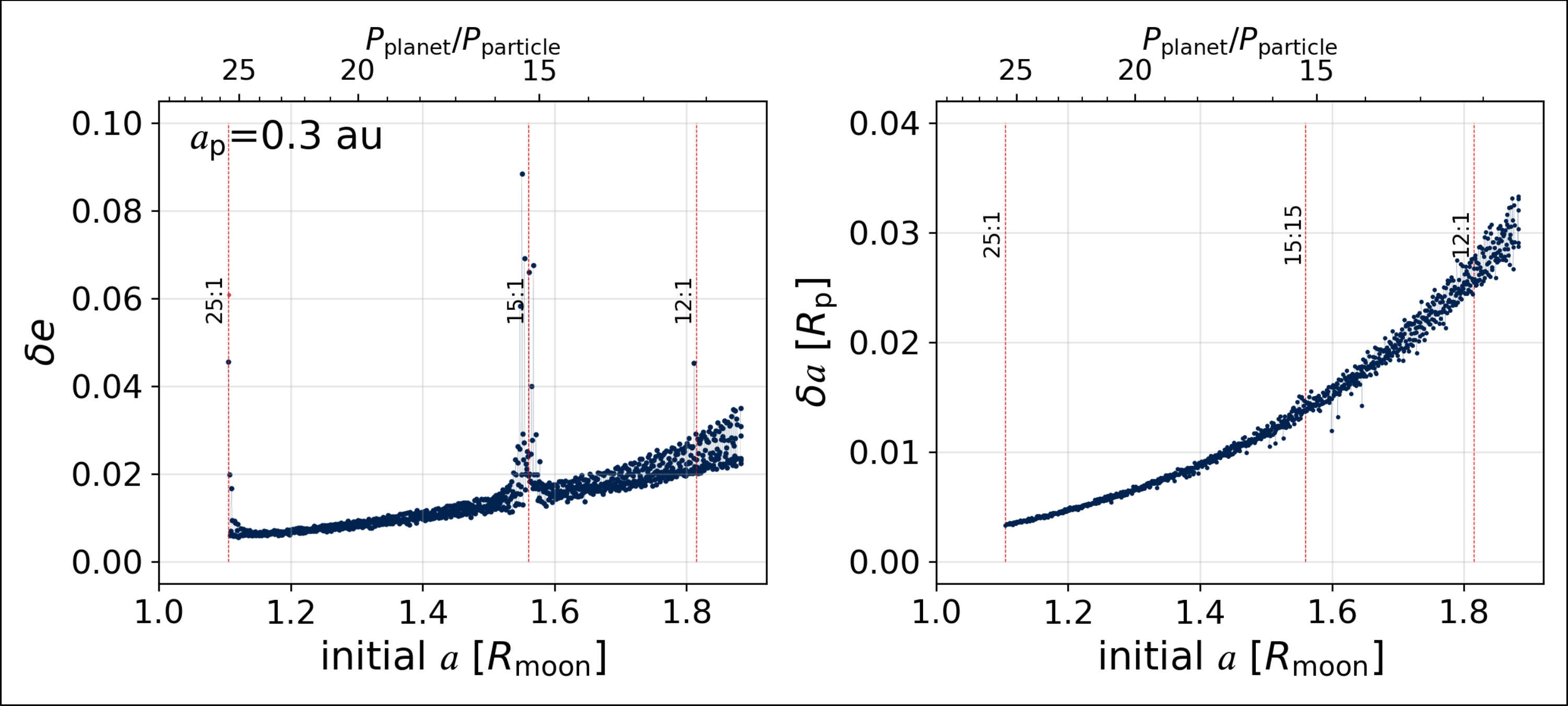}
        	\hspace{0.15cm}
        	\vspace{0.25cm}
        \includegraphics[scale=0.33]{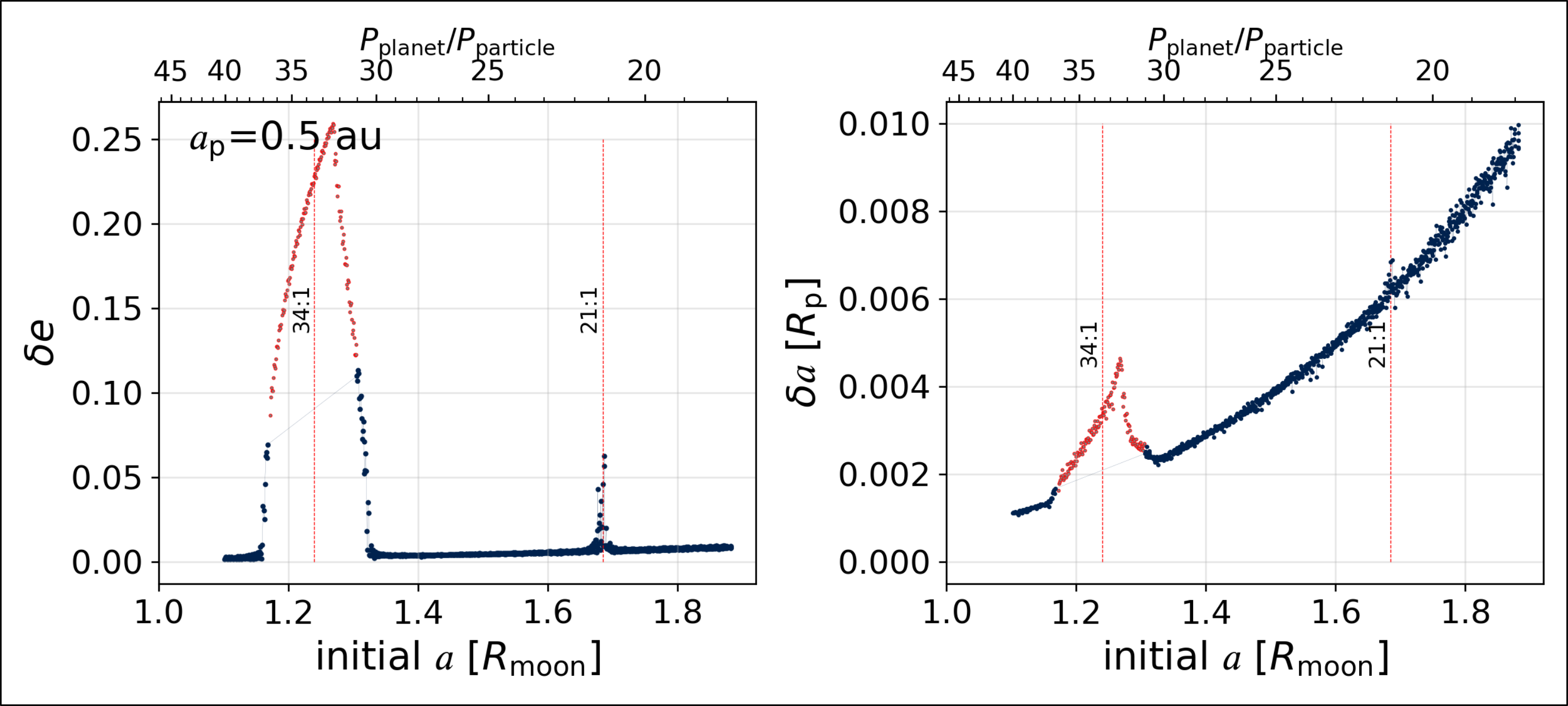}
     \end{subfigure}
          \hspace{0.15cm}
     \begin{subfigure}[b]{\textwidth}
        \centering
            \hspace{0.15cm}
        \includegraphics[scale=0.33]{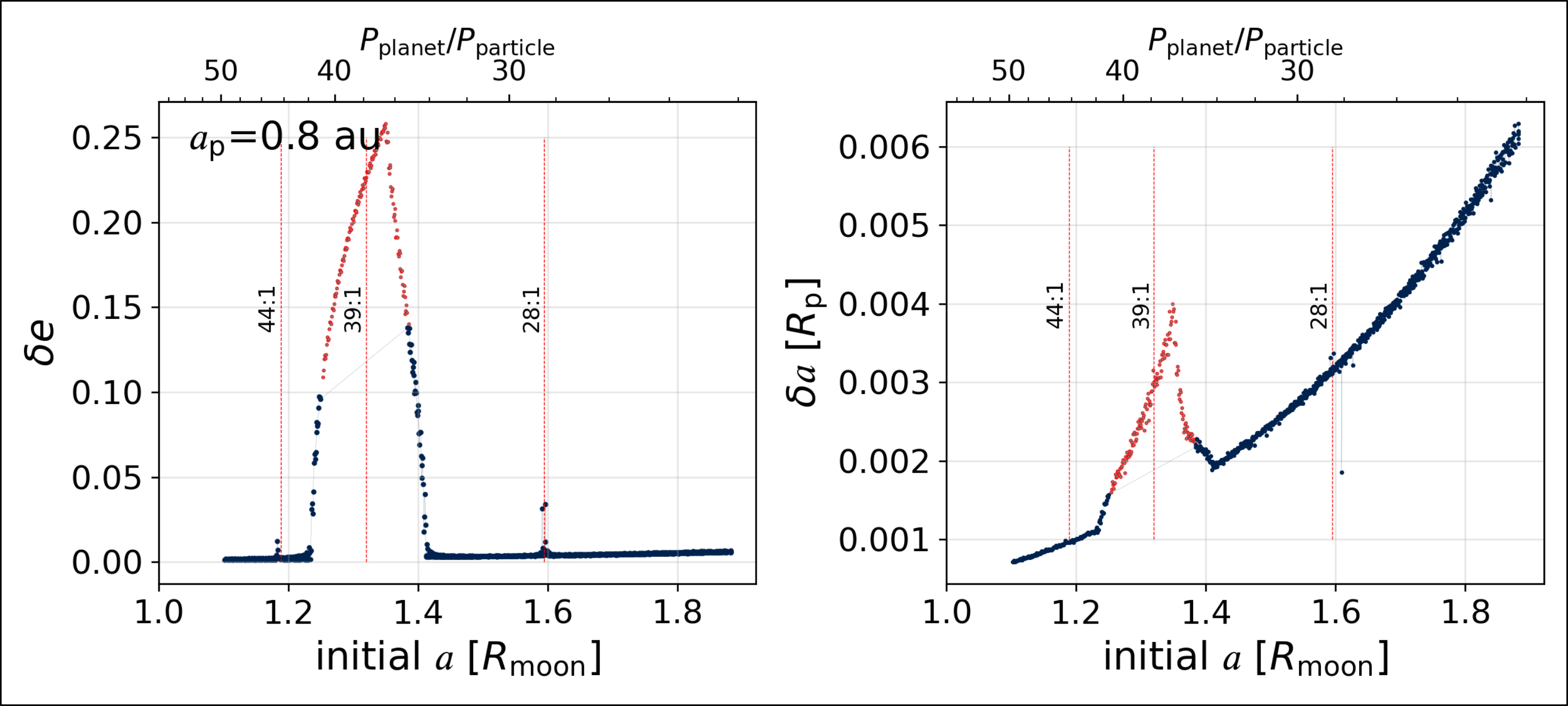}
        	\hspace{0.15cm}
        \includegraphics[scale=0.33]{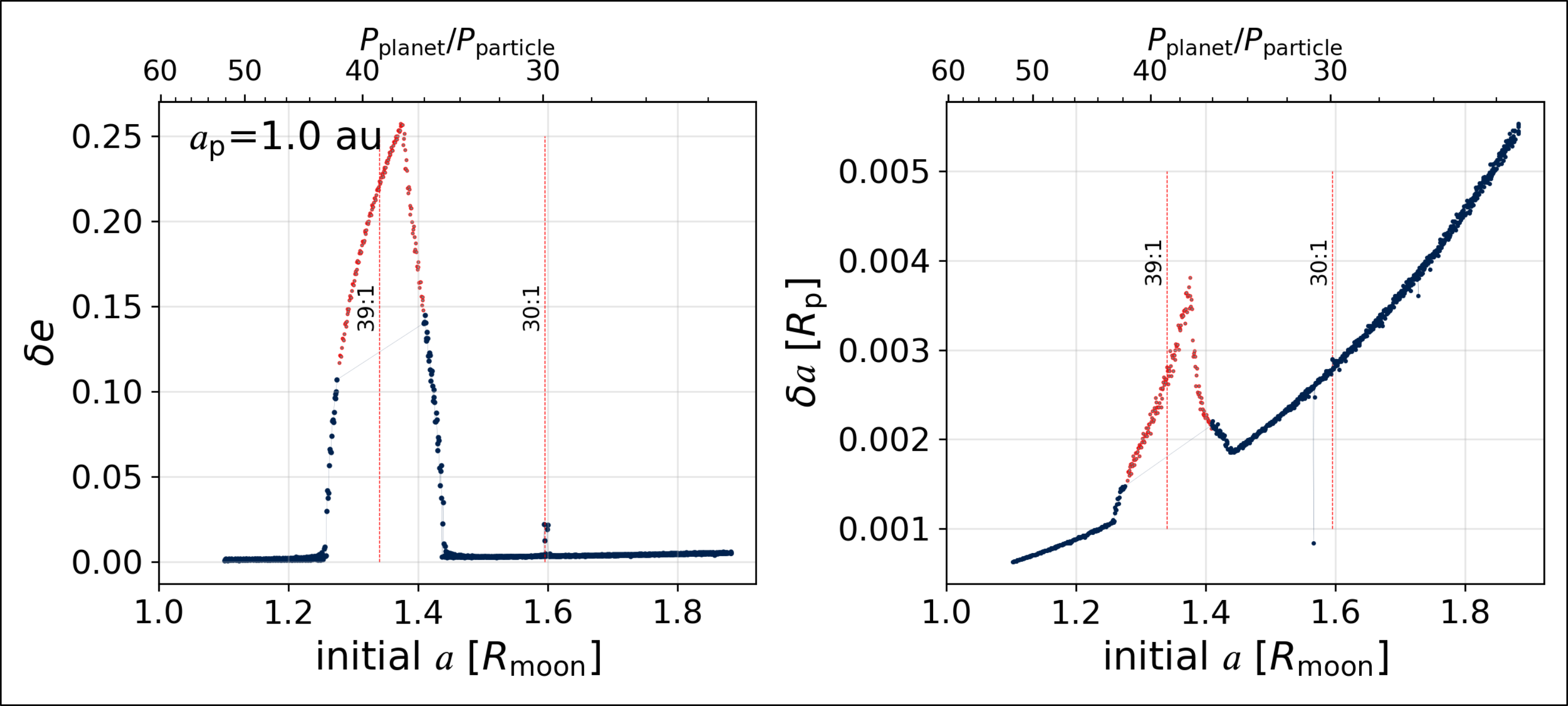}
        %\caption{$\ap = 0.5$ au, $i=0$ deg. }
     \end{subfigure}
     \hfill
     \caption{Same as Fig. \ref{fig:ringsi0} but for a rings' tilt of $30^{\circ}$.}
    \label{fig:ringsi30}
\end{figure*}

 %##################  sub SECTION ########################
\subsection{\textit{Dynamical stability}: rings' size}

After orbital migration of the moon due to planet-moon tidal interactions has already occurred, we adopt the model in \citet{Alvarado2017} to find the corresponding $\astop$ \citep{Sucerquia2020} for a given set of $\ap$, as well as planet and moon's masses, $\Mp$ and $\Mmoon$ (see Section \ref{sec:tidal} and Fig.~\ref{fig:constraints-am}). Once the parent moon is fixed at the aforementioned orbital position, the dynamical stability of its rings' particles is studied by integrating the equations of motion of a set of test-particles evolving under the gravitational influence of the moon, the planet, and the star. This allows us to constrain the inner and outer rings' radii, rings' stability, and lifespan.
% \vspace{-0.3cm}

To set up the initial conditions of the rings' particles (i.e. positions and velocities) we used $r_\mathrm{Roche}$ as the maximum outer limit of the rings,  assuming that both the planet and the satellite are completely spherical and their orbits are co-planar. Under these assumptions, the ring-systems maintain their shape as long as a significant depletion of particles via thermodynamic processes is averted (see previous Section). Also, the initial positions of the rings' particles were generated randomly by implementing blue-noise sampling algorithms\footnote{We use the {\tt fibpy} algorithm (https://github.com/matt77hias/fibpy), which is based on the work of \cite{VOGEL197}.}, in order to avoid oversampling problems. 

We prepared two sets of simulations: 1) {\it icy rings} having particles with densities similar to water ice, $\rho = 0.92$ g cm$^{-3}$ (left-hand panel of Fig.~\ref{fig:comparison}); and 2) {\it dusty rings} composed of refractory particles with densities similar to dust, $\rho = 1.50$ g cm$^{-3}$ (right-panel of Fig.~\ref{fig:comparison}). The host is a Titan-sized moon with a density of $\rho_\mathrm{m} = 3.00$ g cm$^{-3}$, which orbits a Saturn-like planet. Both sets of rings were divided into two subsets according to the rings' inclination, either $0^{\circ}$ or $39^{\circ}$. These angles were arbitrarily selected to assess the effect of the inclination on the orbital evolution of rings' particles, while avoiding values greater than 39.2º where Kozai--Lidov resonances appear \citep{kozai59,Lidov62} and produce evident effects in orbiting ringed structures (see e.g. \citealt{Sucerquia2017}). For each subset, we set the initial position of the planet at $\ap=$ 0.3, 0.5, 0.8, and 1.0 au.

The motivation to include icy rings rises from the particles' lifespan calculations presented in Fig.~\ref{fig:timescale-pr} and the possible formation mechanisms that were explained in Section \ref{sec:scenarios}. The tangible difference between the simulations of icy and dusty rings is their initial size which is calculated via equation \ref{eq:roche}. As it can be seen in Fig.~\ref{fig:comparison}, icy rings are much richer dynamically and present more regions of resonance, while dusty rings are more stable and present a lower scattering of the constituent particles. The systems were allowed to evolve over $\sim 750.000$ periods of the rings' particles, using the N-body integrator {\tt REBOUND} with a {\tt leapfrog} approach \citep{rebound} \rev{and a time-step of 10$^{-5}$ yr}. Figs \ref{fig:ringsi0} and \ref{fig:ringsi30} show the outcomes of the simulations for inclinations of $0$° and $30$°, respectively.

The upper panels of Fig.~\ref{fig:comparison} show a snapshot of the simulations after 100 yr of evolution, where particles that collided with the satellite are marked with red and eccentric particles are depicted in blue. The lower panels of Fig.~\ref{fig:comparison} (as well as Figs \ref{fig:ringsi0} and \ref{fig:ringsi30}) present the semi-amplitude of the excitation in the eccentricity ($\delta e$) and semi-major axis ($\delta a$) experienced by the rings' particles due to the dynamical perturbations produced by the planet and the star. For the case illustrated in Fig.~\ref{fig:comparison} we used 10.000 particles, while the systems presented in Figs \ref{fig:ringsi0} and \ref{fig:ringsi30} were composed of 1000 test particles.

Under the previously mentioned conditions, we found that ringed moons are stable structures with regions that are strongly affected by the periodic perturbations of the planet and the star. Increasing the eccentricity or modifying the semi-major axes of the particles favoured the creation of gaps in the rings. \rev{For example, in the system with $\ap$ = 0.3 au there are at least three notorious regions of mean motion resonances (MMR) with the planetary semi-major axis (i.e. 1:15, 1:20 and probably 2:25 MMR in the lower panels of Fig.~\ref{fig:comparison})}. 
\rev{In Figs \ref{fig:ringsi0} and \ref{fig:ringsi30} we identify the location of the most important $N/1$ MMR between particles and the planet's position that roughly coincides with some spikes. We think that the additional spikes observed in $\delta_a$ and $\delta_e$ could be a consequence of higher order MMR (e.g, $N/2$, $N/3$, etc) and secular resonances that could be present within this region (e.g. the $\nu6$ resonance due to the star and mutual inclinations between the bodies, see \citealt{Murray2000}). However, a thoughtful analysis of the resonances is beyond the scope of this paper and we leave it for a forthcoming work.}

Particles that were affected by resonant perturbations increased their eccentricity, becoming prone to collide with surrounding particles of the system. This, in addition, could increase the particles' orbital decay rate and thus the size of resonance gaps.
Nevertheless, we did not include the effect of such collisions in our simulations, so the gaps did not increase their size in a prominent fashion. Additionally, we have shown here that particles remained fairly well-confined inside the Roche boundary of the system and, therefore, it is reasonable to think that their assumed initial size was an optimal outer radius. This assumption is used in the next section to establish the detectability of ringed moons.

\vspace{-0.6cm}

%#####################################
\section{Effects in light curves}
\label{sec:detectability}

\textit{Cronomoons} have effects on the light curves of transiting exoplanets that are similar to those produced by single moons, but adding special and unique features. For example, they can modify the relationship between depth, duration, and timing of the transit by miscalculations of some physical parameters of the satellite, the planet, and the star.

\rev{Moons orbiting around planets produce TTV and TDV effects on planetary light curves \citep*{Kipping2009c}. In the case of massive moons, these would push their parent planet further away from their mutual centre of mass than less massive moons do}, while also changing more significantly the planet's position and velocity. However, if the actual size of the moon is overestimated (see \citealt{Zuluaga2015}), TTVs and TDVs could be imperceptible. An example of this is a low-mass moon with a ring system that makes it look much larger than it really is (see Section \ref{sec:k1625b}).

\subsection{Transit depth and TRV of ringed exomoons} 
\label{sec:transdepth}

Owing to their extended rings-system, some transiting exoplanets can exhibit apparent sizes that exceed their actual size. This is due to the fact that the transit technique only cares about the portion of the stellar disc that is obscured during a transit; thus, larger areas would immediately mean bigger planets. However, ringed planets can be smaller than what is predicted from their transit properties (see \citealt{Zuluaga2015}). In those cases, the fractional change in stellar flux or transit depth, given by $\delta = (F_\star - F)/F_\star$, mainly depends on the tilt or orbital inclination of the rings towards the observer ($i_\mathrm{R}$), the inner and outer radii of the rings ($f_\mathrm{i}$ and $f_\mathrm{o}$ respectively, in units of $\Rp$); and the opacity of the rings ($\tau$) defined in terms of the light that they absorb \citep{Barnes2004}.

Assuming a spherical moon and a constant optical opacity for its rings-system, the transit depth can be estimated as follows \citep{Zuluaga2015},
\begin{equation}
\label{eq:delta}
\delta = \frac{A_\mathrm{Rp}}{A_\star},
\end{equation}
where $A_\star$ is the area of the stellar disc and
\begin{equation}
\label{eq:aproj}
    A_\mathrm{Rp} \;= \; \pi \Rp^2 + \pi \left[r^2(f_\mathrm{o}) - r^2(f_\mathrm{i})\right]
\end{equation}
is the effective area projected by the planet and its rings. $r^2(f)$ stands for the effective rings' radius projected towards the star, and it is a function of the rings' size and opacity. For more details on $r^2(f)$ see Eq. 3 in \citep{Zuluaga2015}.

\begin{figure}
	\includegraphics[width=\columnwidth]{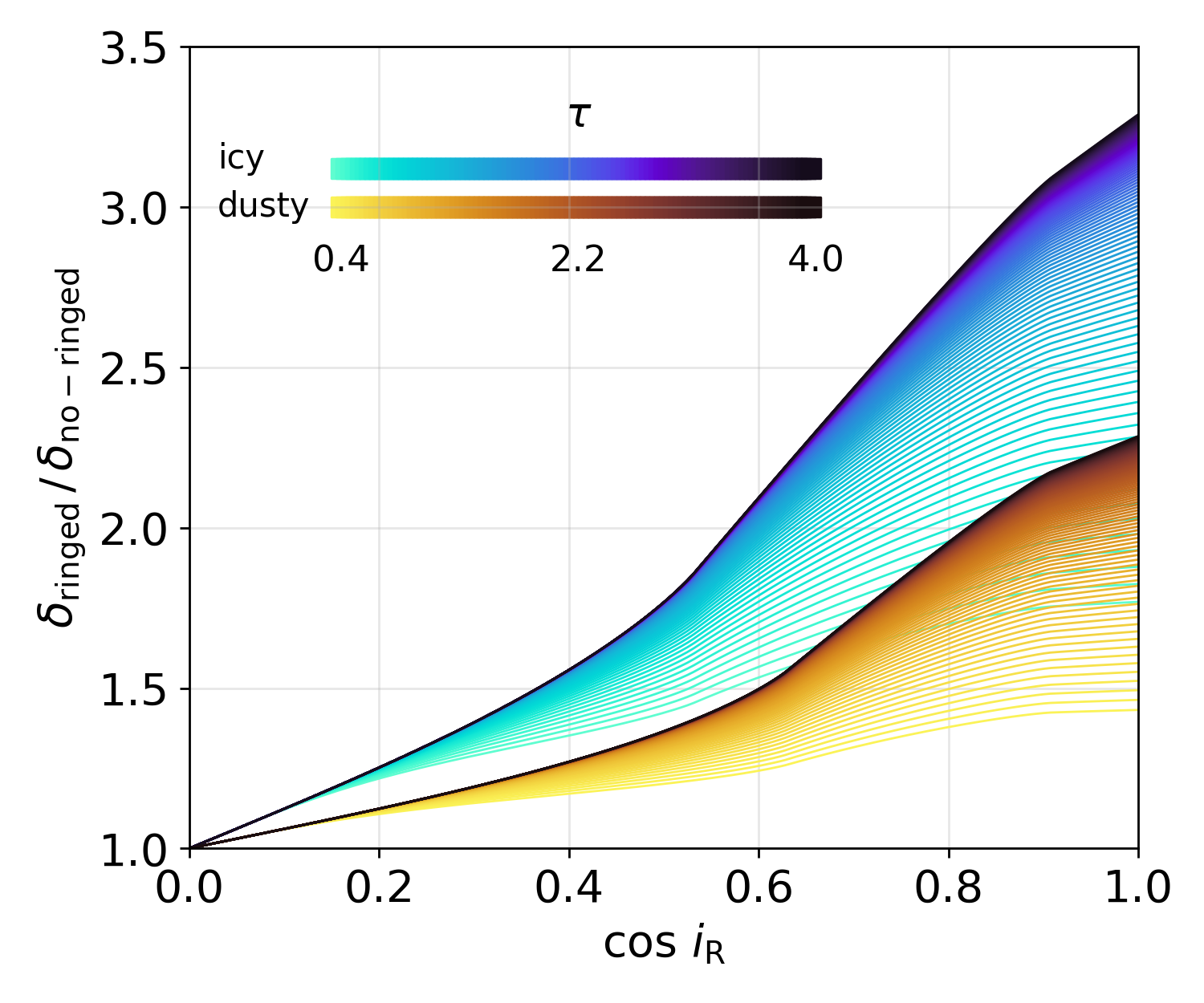}
	\includegraphics[width=\columnwidth]{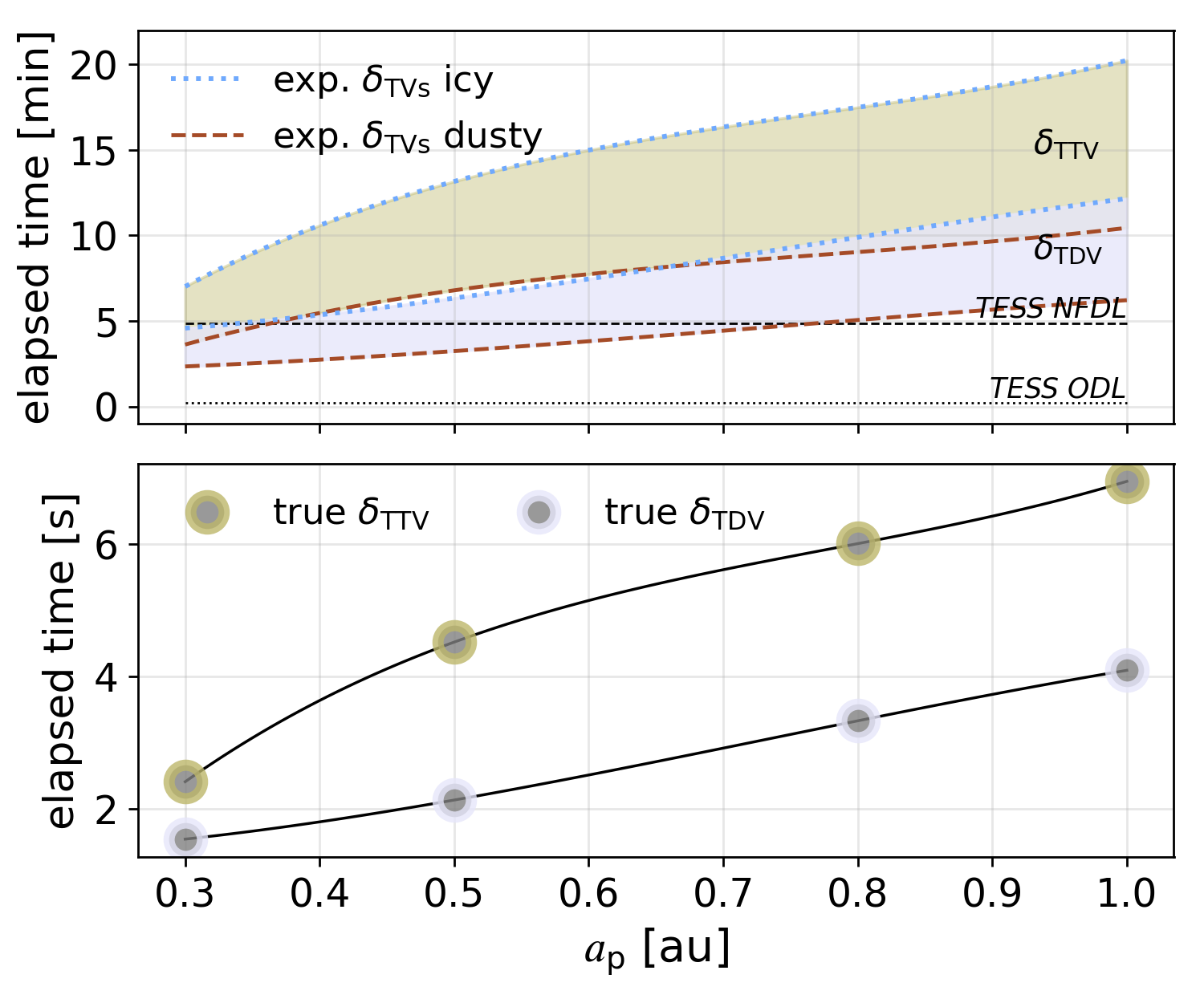}
    \caption{Upper panel: Fraction between the observed and true transit depth for a ringed exomoon as a function of the projected rings' tilt ($\cos \iR$) over the observer line of sight, and different values of the rings' normal opacity ($\tau$) for icy and dusty rings.
    Lower panel: rms amplitudes of TTV (orange) and TDV (blue) signals for the system studied in section \ref{sec:stability_dynam}. The coloured dots are the {\it true} values obtained from the moon's mass and size used in the aforementioned simulation, while the shadows are the expected values for massive moons. \rev{For reference, we include the observational limits for {\it TESS}. That is, the \textit{Near Faint Detection Limit} (black-dashed line) which corresponds to a faint star with a small planet, and the \textit{Optimal Detection Limit} (black-dotted line) related to a star 10x brighter with a planet 2.5x larger than the previous case \citep{Deeg2017}.}}
    \label{fig:transits}
\end{figure}

By applying Equations \ref{eq:delta} and \ref{eq:aproj} to the rings' sizes estimated in the previous section; and for several orientations and compositions of the system (see the upper panel of Fig. \ref{fig:transits}), we found that ringed moons crossing the stellar disc during the \textit{flat} part of the planet's transit will induce flux changes from $\sim 1.5$ to $3.2$ times larger than non-ringed moons. For a system composed of a solar-type star and a Saturn-like planet with an Titan-like moon, the planet and moon would induce a drop in stellar flux of $\delta F_\mathrm{p} \sim 7000$ ppm and $\delta F_\mathrm{m}\sim14$ ppm, respectively. Over different transit measurements, this could potentially lead to apparent Transit Radius Variations \citep{Rodenbeck2020} of about $0.21$ per cent. In the scenario of a ringed moon, the stellar flux variation associated solely to the moon will be up to $\delta F_\mathrm{m-ringed}\sim33$ and $\sim47$ ppm for dusty and icy rings, respectively. Such a system would have a variation in the transit radius of $\sim 0.47$ (dusty) and $\sim 0.77$ (icy) per cent over consecutive periods. Also, moons with highly inclined orbits could induce variations on the inferred transit's impact parameter over consecutive eclipses.

\subsection{TTV and TDV of ringed exomoons} 

A planet with its massive satellite will orbit a common centre of mass, and the mean orbital advance of the planet will be given by the orbital advance of the centre of mass plus/minus the planet's velocity relative to the centre of mass. A distant observer can notice that the planet may either delay or anticipate its transit due to changes in the position of the planet's orbit caused by its moon, producing the so-called Transit Timing Variation (or TTV). Also, during a transit event any change in the velocity of the planet can induce a shortening or widening of the transit duration during different eclipses, something also known as Transit Duration Variation (or TDV). The optimal TTVs occur when orbits are circular and co-planar for both the planet and the moon. For edge-on orbits and from the line of view of an observer on Earth, the barycentric root-mean-squared amplitude for TTVs ($\delta_\mathrm{TTV}$) is given as follows,
\begin{equation}
\label{eq:dttv}
    \delta_\mathrm{TTV} =  \frac{1}{2 \pi} \frac{ \astop \,\Mm}{\ap \, \Mp} \Pp \sqrt{\frac{\Phi_\mathrm{TTV}}{2 \pi}},
\end{equation}
where $\Phi_\mathrm{TTV} \rightarrow \pi$ is the scaling factor for co-aligned circular orbits, as detailed in \citet{Kipping2009b}. In a similar way, the optimal rms velocity-induced $\delta_\mathrm{TDV}$ for the aforementioned configuration reads
\begin{equation}
\label{eq:dtdv}
    \delta_\mathrm{TDV} = \bar{\tau}  \frac{ \astop \,\Mm \Pp}{\ap \, \Mp \; \Pm} \sqrt{\frac{\Phi_\mathrm{TDV}}{2 \pi}},
\end{equation}
with $\Pm$ the orbital period of the moon and $\bar{\tau}$ the averaged transit duration given by: 
\begin{equation}
    \rev{\bar{\tau} = \frac{\Pp}{\pi} \arcsin{\left[\sqrt{\frac{(R_* + \Rp)^2 - (b\;R_*)^2}{\ap^2 - b^2 R_*^2 }}\right]}},
\end{equation}
where $R_*$ is the stellar radius. We have assumed here the most favourable case for the impact parameter (i.e. $b=0$). Also, it is worth noting that Equations \ref{eq:dttv} and \ref{eq:dtdv} do not depend on the moon's actual or apparent size; hence, the rings-system (whose mass is much smaller than that of the moon) has a negligible effect on the amplitude of TTVs and TDVs \rev{for a given semi-major axis}. However, a deep transit (such as that generated by a ringed moon) is expected to be associated with a massive moon and to generate large $\delta_\mathrm{TTV}$ and $\delta_\mathrm{TDV}$ amplitudes which are directly proportional to the moon's mass. 

\rev{Although there exists a degeneracy as evidenced from equations \ref{eq:dttv} and \ref{eq:dtdv}, the ratio of TDV/TTV signals (or TVs for short) can allow us to recover the mass of moons. To predict TVs after a single observation it is also possible to estimate the mass} of moons through a simple mass-radius scaling-law assuming a bulk composition and density for the moon. By following \citet{Zeng2019}, ringed moons with radii inferred from the transit light curves introduced in Section \ref{sec:transdepth} and fully made of refractory material (MgSiO$_3$) will have estimated masses of $4.17$ and $8.08$ times the Earth's mass (for dusty and icy rings, respectively), producing significant effects on TVs as shown by the middle and lower panels of Fig. \ref{fig:transits}. There, expected TVs induced by massive moons (denoted as exp. in the plots) are of the order of minutes and are depicted with light green/blue for TTVs/TDVs. The lower limit of each region corresponds to rings made of dust (brown dashed lines), whereas the upper limit represents rings made of ices (blue dotted lines). For the parameters adopted in our simulations (see Section \ref{sec:stability_dynam}), the lower panel of Fig.~\ref{fig:transits} shows the TVs labelled as `true' (dot markers). \rev{Finally, in the same figure we include the \textit{Near Faint Detection Limit} (NFDL; 292 s) for the Transiting Exoplanet Survey Satellite ({\it TESS}), which corresponds to a faint a star with a small planet, as well as the \textit{Optimal Detection Limit} (ODL; 12 s) for a star 10x brighter with a planet 2.5x larger. For a full description of the parameters involved in each scenario the reader can refer to table 1 in  \citet{Deeg2017}.}

To summarise this section, \textit{cronomoons} can produce deep imprints on the stellar flux (i.e large TRVs) but at the same time \rev{small/undetectable} transit secondary effects such as TTVs and TDVs. This is a result of having icy or dusty rings-systems around them.

%%%%%%%%%%%% SECTION %%%%%%%%%%%%%%%%
\section{The system Kepler-1625b i}
\label{sec:k1625b}

We studied the system Kepler-1625b i under the physical scenario proposed in this work. \rev{Despite Kepler-1625b i is the most accepted exomoon candidate whose dip was first reported by \citet{Teachey-Kipping-2018}, there is still no consensus to date on the existence of such an exomoon nor its physical/orbital properties. For instance, \citet{Kreidberg2019} disagrees with the existence of the exomoon's dip, although their work has been shown to be afflicted by higher systematic noise issues \citep{Teachey_2020}. Still, there exists some convergence: \citet{Rodenbeck2020} independently recovered the dip and the three teams have analysed the same data and agree with the TTV evidence}. Kepler-1625b i is thought to be a massive Neptune-sized moon \citep*{Teachey-etal-2018}, but this moon might well be a less massive satellite surrounded by an optically thick rings-system whose dynamical features could be inferred by following the formalism developed in the previous Sections.

If we assume that a moon has actually been detected through a second stellar flux drop in the light curve of Kepler-1625b i, its depth would allow us to deduce the moon's size $\Rmo$ and consequently infer the moon's mass $\Mmo$ through scaling laws (e.g. that of \citealt{Chen2016}). In addition, the moon's semi-major axis can also be constrained from the duration of the dip produced by the moon. Table \ref{tab:properties} summarises some attempts to calculate the physical and orbital properties of the planet and the moon. Then, complementary methods to transit photometry can be used to fully characterise the system, and after consecutive transits the signals associated with secondary effects should conform to those predicted by Equations \ref{eq:dttv} and \ref{eq:dtdv}. If that is not the case the predicted physical parameters must be reviewed. In particular, if large TRVs are found along with small TTVs and TDVs, we could be in the case described in the previous sections. What would these signals be if Kepler-1625b i were a \textit{cronomoon}?

\begin{table}
\centering
\begin{tabular}{cccccccc}\hline \hline

 $M_\mathrm{star}$ & $R_\mathrm{star}$ & $M_\mathrm{p}$ & $R_\mathrm{p}$ & $a_\mathrm{p}$ & $\Mmo$ & $\Rmo$ & $\amo$			\\
 $M_{\odot}$ & $R_{\odot}$ & $M_\mathrm{Jup}$ & $R_\mathrm{Jup}$ & au & $M_\mathrm{Nep}$ & $R_\mathrm{Nep}$ & $R_\mathrm{p}$  	\\ \hline	
 $1.079$ & $1.793$ & $10.0$ & $1.18$ & $0.87$  & $1.0$  & $1.0$     & $ 40.0$           \\ \hline 
 \multicolumn{5}{c}{This work, with $\iR \sim 90$°} \\
 \hline
 $1.079$ & $1.793$ & $10.0$ & $1.18$ & $0.87$  & $0.18$  & $0.38$     & $ 40.0$           \\
 \hline \hline
\end{tabular}
\caption[Properties]{Physical and orbital properties of the system Kepler-1625 as given by \citet{Morton-etal-2016}, \citet{Mathur-etal-2017}, \citet{Teachey-etal-2018} and \citet{Teachey-Kipping-2018}. Properties compiled by \citet{Moraes_2020} as input for their simulations.}
\label{tab:properties}
\end{table}

Our first consideration is that the observed moon's size $\Rmo$\footnote{Hereafter we will use the notation \textit{observed} (subindex o), \textit{true} (subindex t), and \textit{expected} (subindex e), to refer to the observed parameters, the true parameters (i.e. those deduced after applying the method presented in this paper), and the expected parameters from extra measurements like TVs (see Section \ref{sec:detectability}) obtained with the \textit{observed} values, respectively.} does not correspond to the actual size of the moon but to its Roche limit (or a projection of it onto the observer's line of view), which is the outer radius of the rings-system. Also, we have assumed an edge-on projection of the rings towards the star (i.e. $\iR \sim 90$°, see Section \ref{sec:transdepth}), so from Equation \ref{eq:roche} the moon's \textit{true} size ($\Rmt$) reads, 
\begin{equation}
    \Rmt =\Rmo \left( \frac{\rho_\mathrm{part}}{2\;\rho_\mathrm{m}}\right)^{\frac{1}{3}}.
    \label{eq:realrm}
\end{equation}
Note that for a non edge-on transit, the \textit{true} size of the moon must be obtained by inverting Equation \ref{eq:aproj} instead of Equation \ref{eq:roche}.

Equation \ref{eq:realrm} will vary according to the composition ($\rho_\mathrm{m}$) chosen for the moon. However, if regular satellites have cumulative masses ($\mathcal{O} [10^{-4}]$, \citealt{Canup2006}) these will range between $\sim 0.31$ and $3.14$ the Earth's mass, corresponding to a moon's density between the Earth's density and that of newfound Super-Earths (i.e. from 5 to 8 g cm$^{-3}$). Therefore, according to Equation \ref{eq:realrm} the minimum satellite radius is about $0.38$ times the radius of Neptune which equates to a moon's \textit{true} mass of $\sim3.14$ times the Earth's mass, dubbed here as $\pi$-Earth. That being said, within the framework of this work the exomoon candidate Kepler-1625b i could be a $\pi$-Earth accompanied by an optically thick rings-system made of ices and extending to distances similar in size to Neptune.

The rms values of transit secondary effects ($\delta_\mathrm{TTV}$ and $\delta_\mathrm{TTV}$) can be estimated using Equations \ref{eq:dttv} and \ref{eq:dtdv}; and compared with the values \rev{in \citet{Kipping2021} who computed a TTV rms of 17.08 min.} When using the values mentioned in the paragraph above, we found that $\delta_\mathrm{TTV,e} \sim 9.03$ min and $\delta_\mathrm{TTV,t} \sim 1.67$ min. Moreover, we obtained rms TDVs equal to $\delta_\mathrm{TDV,e} \sim 2.70$ min and $\delta_\mathrm{TDV,t} \sim 0.50$ min. It is worth mentioning that no TDVs have been measured yet, so the difference between expected and obtained values supports the hypothesis of an unseen planetary companion in a resonant orbital configuration (see \citealt{Kreidberg2019,Teachey_2020}).

\begin{figure}
     \centering
        \includegraphics[width=\columnwidth]{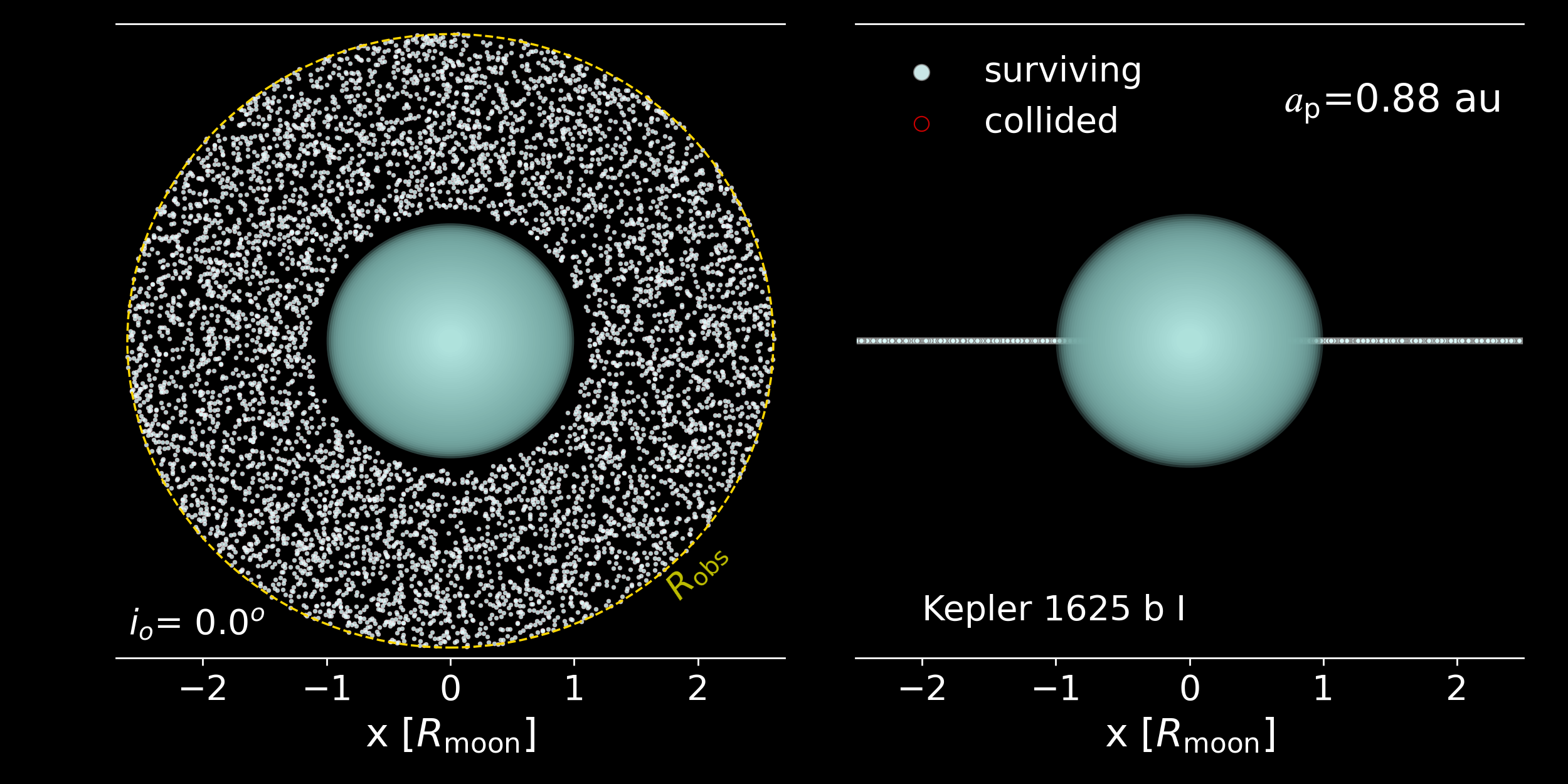}
        \includegraphics[width=\columnwidth]{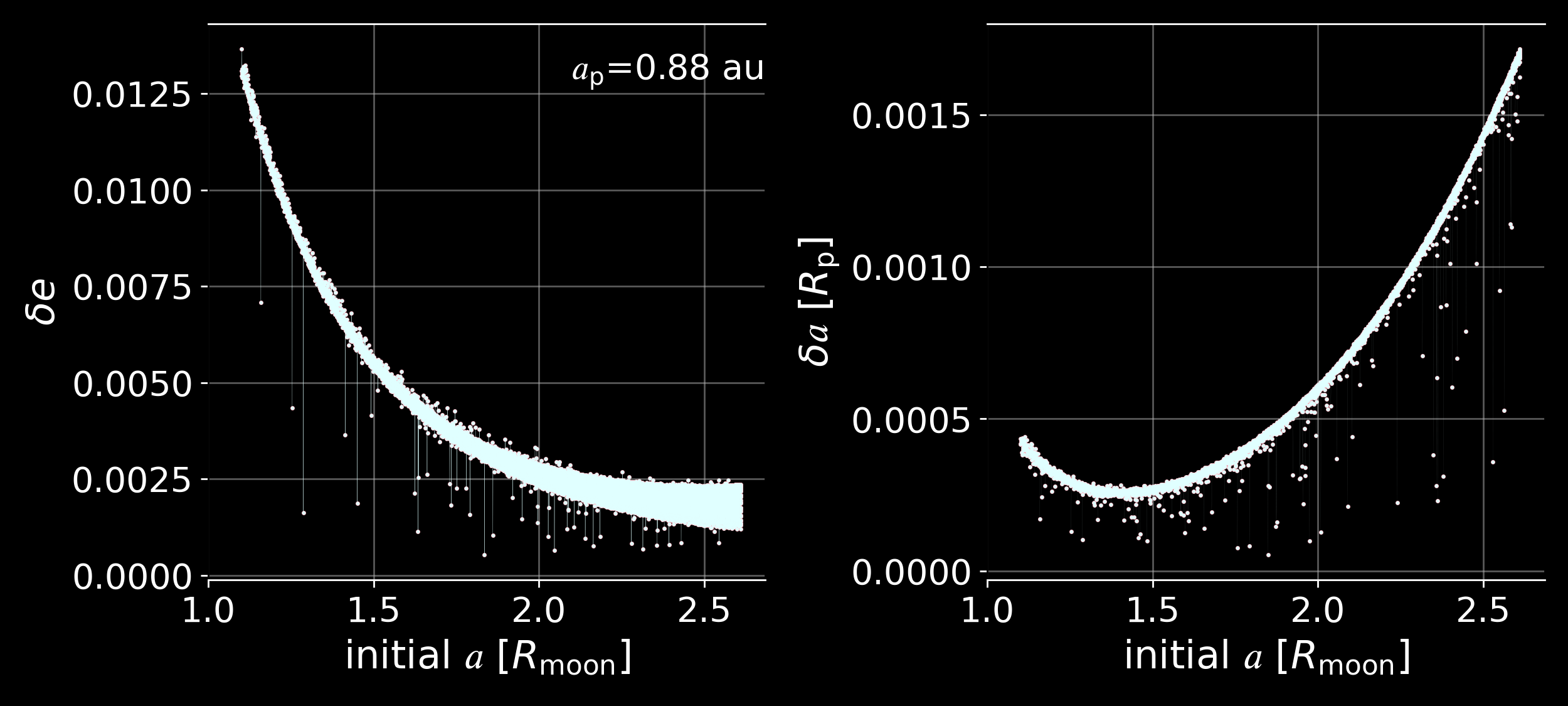}
        \caption{Simulation results for the system Kepler-1625b i assumed as a \textit{\textit{cronomoon}} with icy rings in face--on transit. $R_\mathrm{obs}$ is reported as the observed size of the moon.}
    \label{fig:k6125}
\end{figure}

\begin{figure}
     \centering
          \includegraphics[width=\columnwidth]{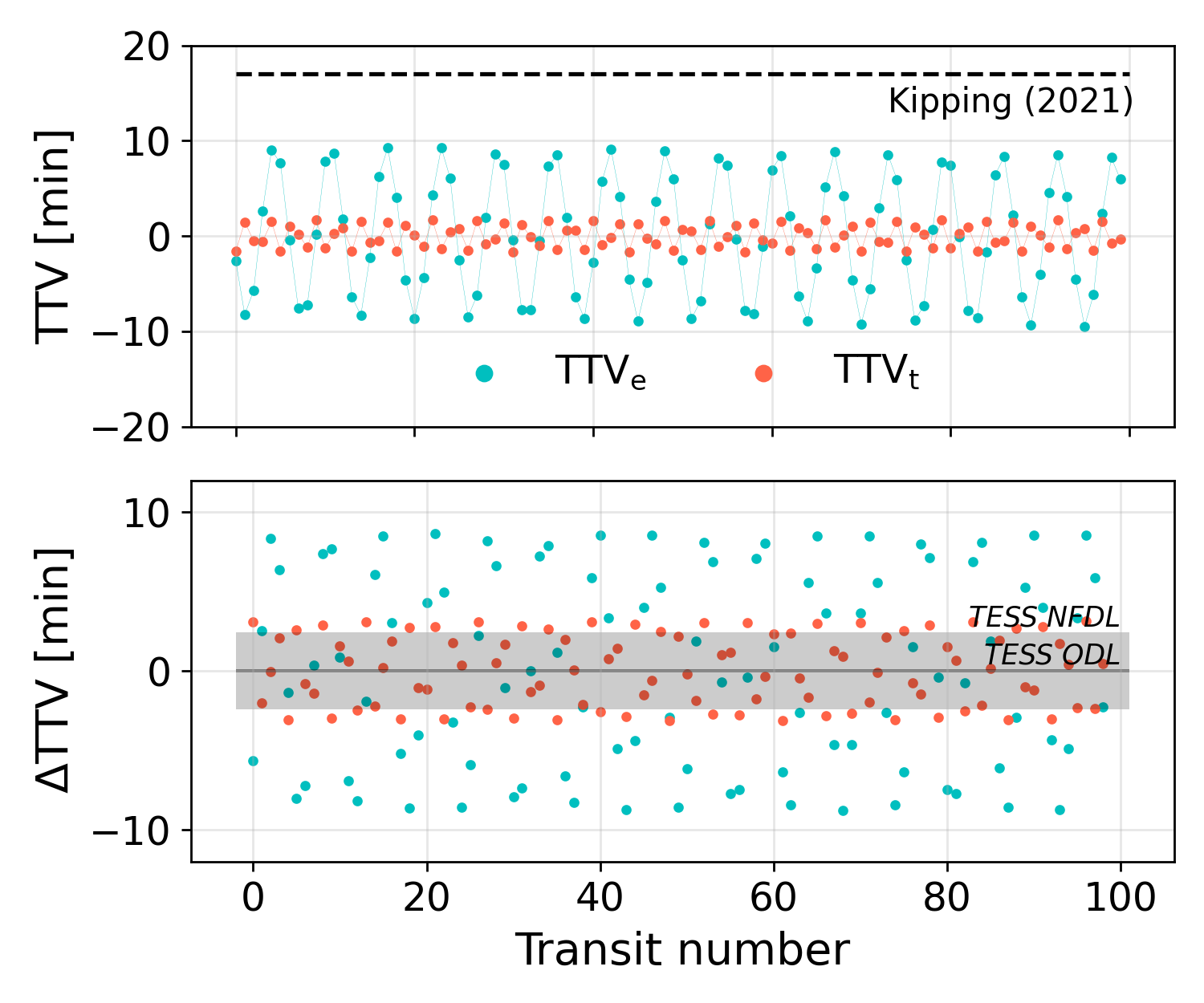}
        \includegraphics[width=\columnwidth]{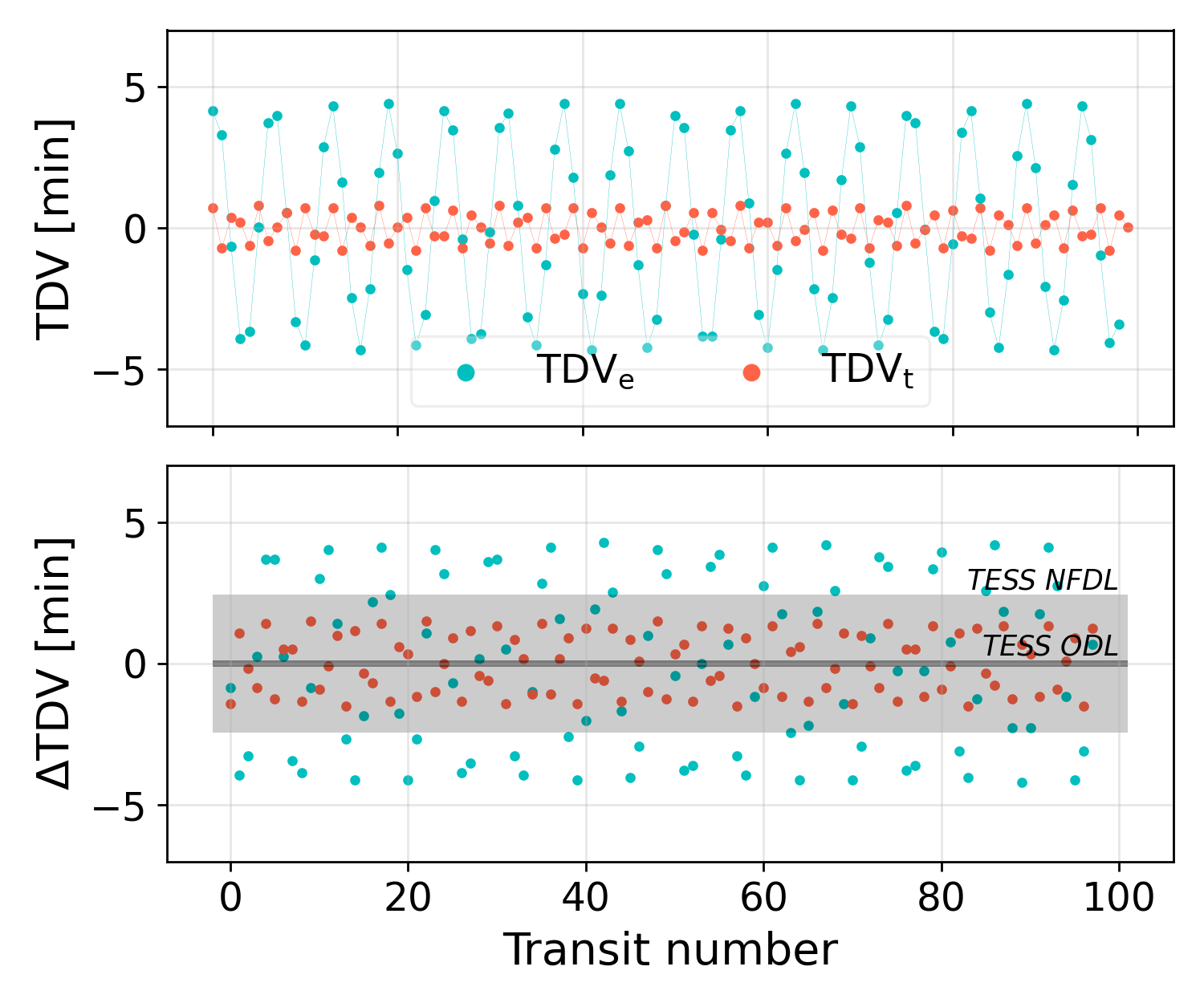}
        \caption{\rev{TTVs and $\Delta$TTVs (upper panel), and TDVs and $\Delta$TDVs (lower panel)} of the system Kepler-1625b i obtained via N-body numerical simulations. Cyan dots represent the values of a massive giant moon, whereas red dots correspond to a {\it cronomoon}. \rev{$\Delta$TTVs and $\Delta$TDVs stand for the elapsed time between consecutive transits and these are compared to the NFDL and ODL for {\it TESS}. Values inside the shaded regions are undetectable.}}
    \label{fig:k6125TTV}
\end{figure}

Adopting the parameters found under the hypothesis of a \textit{cronomoon} listed on Table \ref{tab:properties}, we performed numerical simulations using the N-body code {\tt REBOUND} \citep{rebound} and a 15th order Gauss-Radau integrator (IAS15;  \citealt{reboundias15}). In the absence of any other perturbing body, unlike the cases studied in Section \ref{sec:stability_dynam}, we found that the orbits of the rings' particles were very regular (lower panel in Fig. \ref{fig:k6125}) and did not present any orbital resonances or gaps (upper panel in Fig. \ref{fig:k6125}). Also, we extracted TTVs and TDVs for Kepler-1625b i when orbited separately by a Neptune-like moon and a \textit{\textit{cronomoon}}, represented by the blue and orange dots in Fig. \ref{fig:k6125TTV}, respectively. \rev{Also, the difference in TTVs and TDVs between consecutive transits are shown in Fig. \ref{fig:k6125TTV}, whose values are compared to the NFDL and ODL for {\it TESS}. $\Delta$TTVs and $\Delta$TDVs outside these limits could be detected by {\it TESS}.} We found that both sets of TTV amplitudes are in agreement with those calculated via equation \ref{eq:dttv}, contrary to TDV signals whose values are much larger than expected. The origin of this difference between the measured and expected TDV values lies in the long orbital period of the planet (with its moon) around the common centre of mass, as compared to the transit duration. We computed the period for both the Neptune-like moon and the \textit{\textit{cronomoon}} as 12.6 d, with an expected transit duration of 13.5 h.

\rev{It is important to note that the parameters compiled in Table \ref{tab:properties} were obtained assuming the existence of a moon with no rings. Therefore, a new analysis of the data by assuming a moon with rings could lead to very different conclusions regarding the estimates of mass and semi-major axis of the moon, while reproducing too the deep drop in stellar brightness originally associated with a giant moon.}
 
\begin{figure}
     \centering
          \includegraphics[width=\columnwidth]{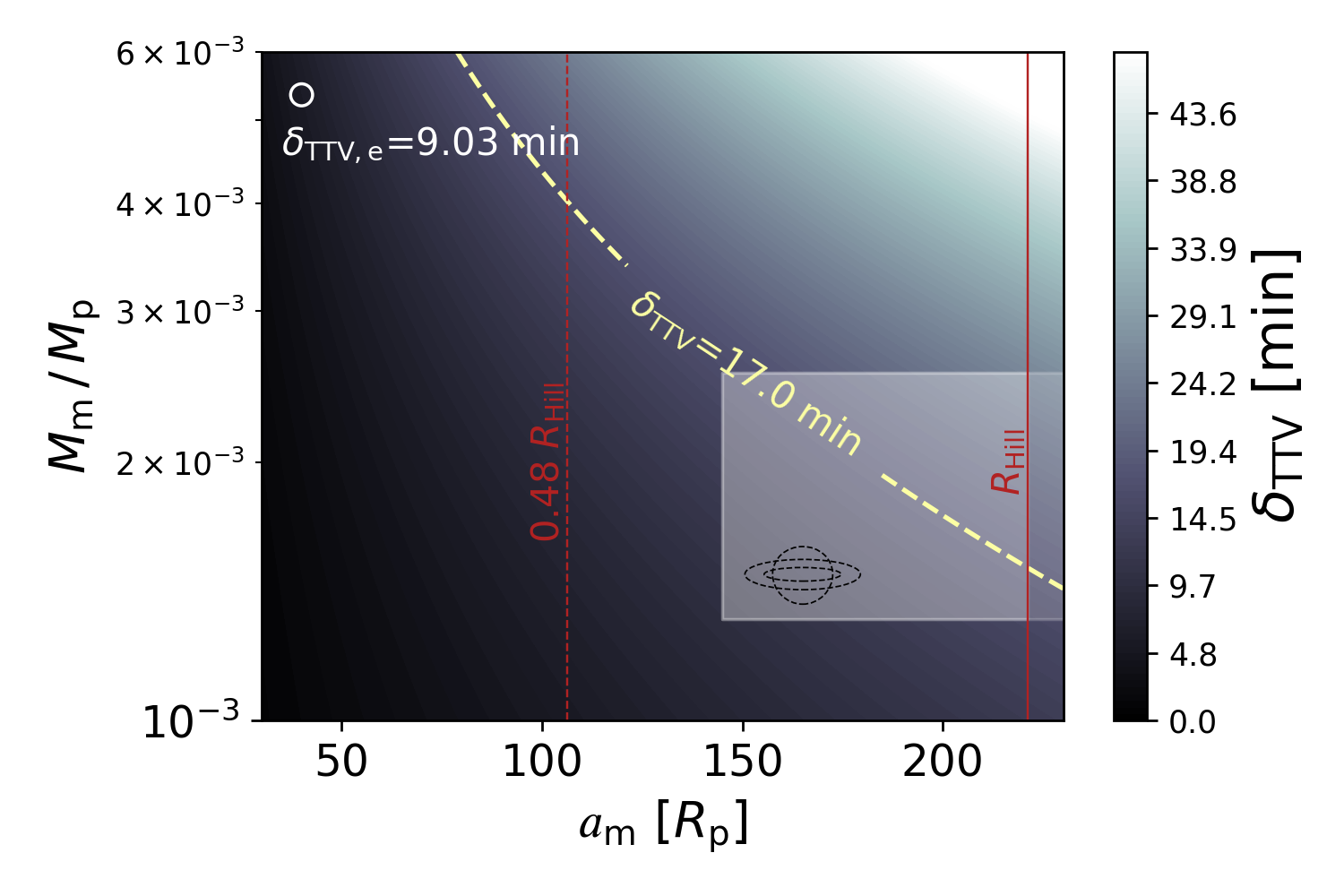}
        \caption{\rev{TTV map (equation \ref{eq:dttv}) for Kepler-1625b i for a range of masses and planet-moon separations. The white circle represents the $\delta_\mathrm{TTV}$ found with data from Table \ref{tab:properties}, while the dashed yellow line presents the mass-distance combinations that account for the TTV obtained by \citet{Kipping2021}. The white square is the region where a \textit{cronomoon} is compatible with such a measurement. Vertical red lines are the satellite stability limits.}}
    \label{fig:chronottvmap}
\end{figure}

\rev{Additionally, if it is assumed in such calculations the 17-min TTV reported by \citet{Kipping2021} which differs from the 9-min one obtained with the parameters of the previously mentioned table, a less massive moon (or a {\it cronomoon}) than that reported should be located at a much more distant position than previously thought. Although the analysis of the Kepler-1625b i system under the framework of this research including the tidal migration and orbital stability of the satellite will be explored in a future work, Fig. \ref{fig:chronottvmap} shows the orbital separations that reproduce a TTV of 17 min for the range of lunar masses compatible with dust and icy rings. In these circumstances, the satellite should be located between 4 and 5 times the initially reported position of 40 times the planet's size (i.e. the white square in Fig. \ref{fig:chronottvmap}). These new positions are inside the planet's Hill limit ($R_\mathrm{Hill}$), but beyond its secondary Hill limit (i.e. 0.48 $R_\mathrm{Hill}$), so that orbital stability would not be fully guaranteed.}

%-----------------------SECTION------------------------------------
\section{Summary and discussion}
\label{sec:discussion}

We studied the likelihood for ringed satellites around close-in exoplanets to exist and the feasilbity for them to be detectable. We have dubbed these objects as \textit{cronomoons}. These systems could explain some light curves with atypical behaviour (give res to works with atypical behavior).  We explore proportionality between the inferred size of {\it cronomoons} and their dynamical effects on their host planet in the context of Transit Timing Variation (TTV) and Transit Duration Variation (TDV). We predict the TTV and TDV signals would be much smaller than variations in the measured moon's radius (TRV).

To analyse the dynamical stability of \textit{cronomoons} we used satellite migration models to determine their asymptotic semi-major axis, a position where \textit{cronomoons} will remain for most of their lifespan. This allowed us to accurately assess the dynamics of their rings by fixing the gravitational effect of the planet and the star on the rings' particles. Additionally, this provides information that can be used to estimate the effects that moons produce on TTVs and TDVs, while helping reduce the degeneracy between these two kinds of signals. We also studied the life expectancy of the constituent particles of rings when they are subject to sublimation processes and orbital decay, finding that such particles have a sufficient life expectancy that make their discovery feasible by recognising morphological changes on planetary light curves (after successive observations and depending upon mean circumstellar distances).

By using N-body numerical simulations we found that periodic perturbations of both the star and the planet are not significant enough to disrupt {\it cronomoons}. Still, such perturbations can modify their rings and enrich their morphology: the most evident effect is the opening of gaps which arise due to resonant perturbations with the planet. As expected, those scenarios where host moons and planets are located in close-in orbits may potentially be more harmful to rings-systems because of the strength of gravitational interactions. However, if \textit{cronomoons} orbit more distant planets and evolve under the migration model used in this work \citep{Alvarado2017}, they will have more stable orbits so long as they keep within the dynamical stability limit of the system. In contrast, scenarios where the physical parameters of the planet are fixed can result with moons located at any distance \citep{Dobos2021}, including short-period orbits where rings can rapidly be dissipated by perturbing mechanisms.

The viability of \textit{cronomoons} as a physical system strongly depends on the environment where they develop and dwell in, as described in the different formation scenarios of Section \ref{sec:scenarios}. For example, mergers between sibling satellites or close encounters with exocomets could populate exomoons with rings, and even result in different types of circumplanetary and circumsatellital structures that can be classified according to their composition and morphology. Collisions between moons could generate co-planar dusty rings, whereas encounters or flybys with exocomets could generate icy rings in random orbital inclinations (taking as a proxy the orbital properties of the comets in the Solar System\footnote{See \url{https://ssd.jpl.nasa.gov/?sb_elem}.}). Additionally, any deviation from coplanarity between the orbits of moons and their rings could be interpreted as a signpost of a recent tidal disruption event that delivered fresh material around the moon to form the rings, given that dynamical evolution tends to a fast coplanarization of this type of systems \citep{Sucerquia2017}. Interestingly, studying transit secondary effects of systems with tilted configurations would allow us to follow them up and monitor them over very long baselines.

Beyond their dynamical stability, \textit{cronomoons} can significantly alter first- and second-order effects on planetary light curves. We showed that they can exhibit transits with large TRV but dimmed TTV and TDV signals (contrary to what is expected for massive systems). In addition, the transit geometry of \textit{cronomoons} at their ingress and egress stages, as well as their rings' gaps may reveal the presence of these objects, but it was also found that large transit depths generated by circumsatellital rings can be related to moderate masses that do not necessarily perturb the dynamics of host planets. 

\rev{Among the ten current claims for exomoon candidates \citep{Benjaffel2014,Bennet2014,Teachey-Kipping-2018,Oza2019,Kipping2020_6m,Quarles2020,Fox2021}}, we chose to study the system Kepler-1625b i (Section \ref{sec:k1625b}) -- the most convincing exomooon candidate so far -- and compare its properties to those of a stable \textit{cronomoon}. Despite an accurate fitting model for the light curve of this system was not used, we tested its dynamical evolution and detectable features by assuming a face-on transit: this implies that the sum of the rings' tilt and the inclination of the system is $\sim90$°, helping us illustrate another phenomenological explanation of the observations by predicting small values for TTVs and TDVs.

In this work we mainly studied the case of rings around moons which orbit close-in planets, as this kind of systems could be easier to detect. However, giant planets located in farther circumstellar orbits could also bear ringed moons whose existence is possible via different mechanisms like capture of moons or destructive orbital resonances. Events such as the Late Heavy Bombardment (LHB) as detailed by the Nice Model of the Solar System, or the recent impact of the Shoemaker--Levy 9 comet on Jupiter (see e.g.  \citealt{Zahnle1994}) could give rise to fortuitous collisions that are, in fact, frequent in the mature stages of planetary systems. Also, despite impacts of comets against the Earth during the K/T mass extinction event \citep{Siraj_2021} are still debated due to a lack of well-established evidence in the Solar System \citep{Desch2021}, this scenario might be somewhat frequent especially during the early stages of planetary evolution and might lead to the formation of rings around moons.

The fact that no {\it cronomoons} have been detected in the Solar System could mean that their lifespan was shorted due to the dispersion of particles in their orbital plane, produced by the presence of other planets and moons (a scenario we will study in a future work). In addition, the small density of most moons in the Solar System could have produced small cross sectional areas for {\it cronomoon} formation after the passage of an outsider object (e.g. during the LHB event), so that capturing material to form a {\it cronomoon} might have been feasible but also unlikely. On the other hand, {\it cronomoons} formed after moon-moon collision events might be more likely and frequent to form in the early stages of the Solar System, but these objects might have already been depleted via the aforementioned processes.

Adding to the scenarios mentioned in Section \ref{sec:scenarios}, a feasible avenue for the formation of {\it cronomoons} in the current Solar System could be the capture of objects already bearing rings. For instance, {\it cronomoons} might be formed if giant planets trapped ring-bearing bodies such as the Centaurus Chariklo and Chiron, or the trans-Neptunian object Haumea given that their rings-systems survive to the gravitational stresses produced by the capture process. Future geological studies applied on surfaces of moons searching for ancient impacts may detect a disparity between their natural composition and the particles that were deposited on their surfaces by older rings-systems. These may possibly form ridge-like structures in a similar way to the process of ring's particles falling on a planet's atmosphere, evidenced by the Cassini's {\it Grand Finale} phase as described by \citet{waite2018} and references therein.

To summarise our work, we studied a feasible scenario whereby rings around moons might flourish and thrive. We dubbed these objects {\it cronomoons} after their similarity with Cronus (the Greek name for Saturn), and after Chronos (the epitome of time), following the effects on light curves in terms of transit timing and duration. \textit{Cronomoons}, in addition to endowing us with a new interpretation of light curves, give us a glimpse about the exotic extraterrestrial landscapes that could take place around exoplanets, and even, in our own Solar System.

\section*{Acknowledgements}
The authors thank the referee, David Kipping, for the helpful feedback and positive remarks, which allowed us to improve the quality of this research. 
Mario Sucerquia acknowledges support by Agencia Nacional de Investigaci\'on y Desarrollo (ANID) through FONDECYT postdoctoral 3210605. Mario Sucerquia, Amelia Bayo, Johan Olofsson, Jorge Cuadra, and Mat\'ias Montesinos thank ANID – Millennium Science Initiative Program – NCN19\_171. Jaime A. Alvarado-Montes acknowledges funding support from Macquarie University through the International Macquarie University Research Excellence Scholarship (`iMQRES'). This project has received funding from the European Union's Horizon 2020 research and innovation programme under the Marie Sk\l{}odowska-Curie grant agreement Nº 210021. Jorge I. Zuluaga is funded by Vicerrector\'ia de Docencia, UdeA. Cristian Giuppone acknowledge the computational resources from CCAD -- UNC (\href{https://ccad.unc.edu.ar}{https://ccad.unc.edu.ar}), which are part of SNCAD -- MinCyT, República Argentina. The authors of this paper call for respect of fundamental rights in order to achieve equity, social justice, and peace needed in Colombia.\\ 
This work is dedicated to Mart\'in.

\section*{Data Availability}
The data underlying this article will be shared on reasonable request to the corresponding author.

%%%%%%%%%%%%%%%%%%%%%%%%%%%%%%%%%%%%%%%%
%%%%%%%%%% REFERENCES %%%%%%%%%%%%%%%%%%
%%%%%%%%%%%%%%%%%%%%%%%%%%%%%%%%%%%%%%%%
\bibliographystyle{mnras}
\bibliography{references.bib} 
%####################################
%\appendix

% Don't change these lines
\bsp	% typesetting comment
\label{lastpage}
\end{document}